\pdfoutput=1

\documentclass[11pt,twoside,a4paper,cmspaper,final,collab]{cms-tdr}
\def\svnVersion{338039}\def\cmsCernNoTag{CERN-PH-EP/2013-037}\def\cmsCernDate{\today}\def\cmsMessage{Published in the Journal of High Energy Physics as \href{http://dx.doi.org/10.1007/JHEP03(2016)125}{\doi{10.1007/JHEP03(2016)125.}}}

\begin{document}\cmsNoteHeader{EXO-14-015}

\hyphenation{had-ron-i-za-tion}
\hyphenation{cal-or-i-me-ter}
\hyphenation{de-vices}
\RCS$Revision: 337944 $
\RCS$HeadURL: svn+ssh://svn.cern.ch/reps/tdr2/papers/EXO-14-015/trunk/EXO-14-015.tex $
\RCS$Id: EXO-14-015.tex 337944 2016-04-10 19:18:04Z alverson $
\newlength\cmsFigWidth
\ifthenelse{\boolean{cms@external}}{\setlength\cmsFigWidth{0.85\columnwidth}}{\setlength\cmsFigWidth{0.4\textwidth}}
\ifthenelse{\boolean{cms@external}}{\providecommand{\cmsLeft}{top}\xspace}{\providecommand{\cmsLeft}{left}\xspace}
\ifthenelse{\boolean{cms@external}}{\providecommand{\cmsRight}{bottom}\xspace}{\providecommand{\cmsRight}{right}\xspace}

\providecommand\mlstar{\ensuremath{M_{\ell^*}}\xspace }
\providecommand\lstar{\ensuremath{\ell^*}\xspace }
\providecommand\estar{\ensuremath{\Pe^*}\xspace }
\providecommand\mustar{\ensuremath{\Pgm^*}\xspace }
\providecommand\llZ{\ensuremath{\ell \ell \PZ}\xspace}
\providecommand{\Pj}{\ensuremath{\mathrm{j}}\xspace}
\providecommand\llg{\ensuremath{\ell\ell^* \to \ell\ell\gamma}\xspace}
\providecommand\eeg{\ensuremath{ \Pe\Pe^* \to \Pe\Pe\gamma}\xspace}
\providecommand\mmg{\ensuremath{\Pgm\Pgm^* \to \Pgm\Pgm\gamma}\xspace}
\providecommand{\eeee}{\ensuremath{\Pe\Pe^* \to  \Pe\Pe\PZ  \to 4\Pe}\xspace}
\providecommand{\eemm}{\ensuremath{\Pe\Pe^* \to  \Pe\Pe\PZ \to  2\Pe 2\Pgm}\xspace}
\providecommand{\eejj}{\ensuremath{\Pe\Pe^* \to \Pe\Pe\PZ  \to  2\Pe 2\Pj }\xspace}
\providecommand{\mmmm}{\ensuremath{\Pgm\Pgm^* \to \Pgm\Pgm\PZ  \to 4\Pgm}\xspace}
\providecommand{\mmee}{\ensuremath{\Pgm\Pgm^* \to \Pgm\Pgm\PZ  \to 2\Pgm 2\Pe}\xspace}
\providecommand{\mmjj}{\ensuremath{\Pgm\Pgm^* \to \Pgm\Pgm\PZ  \to 2\Pgm 2\Pj}\xspace}
\providecommand{\lljj}{\ensuremath{\ell\ell^* \to \ell\ell\PZ  \to 2\ell 2\Pj }\xspace}
\providecommand{\llll}{\ensuremath{\ell\ell^* \to \ell\ell\PZ  \to 4\ell }\xspace}
\providecommand\Mmin{\ensuremath{M_{\text{min}}}\,  }
\providecommand\Mmax{\ensuremath{M_{\text{max}}}\,  }
\providecommand{\NA}{---\xspace}

\cmsNoteHeader{EXO-12-028}
\title{Search for excited leptons in proton-proton collisions at \texorpdfstring{$\sqrt{s} = 8\TeV$}{sqrt(s) = 8 TeV}}
\date{\today}

\abstract{A search for compositeness of electrons and muons is presented using a  data sample of proton-proton collisions at a center-of-mass energy of $\sqrt{s} = 8\TeV$ collected with the CMS detector at the LHC and corresponding to an integrated luminosity of 19.7\fbinv. Excited leptons ($\ell^*$) produced via contact interactions in conjunction with a standard model lepton are considered, and a search is made for their gauge decay modes. The decays considered are $\ell^* \to \ell \gamma$ and $\ell^* \to \ell \PZ$, which give final states of two leptons and a photon or, depending on the \PZ-boson decay mode, four leptons or two leptons and two jets. The number of events observed in data is consistent with the standard model prediction. Exclusion limits are set on the excited lepton mass, and the compositeness scale $\Lambda$. For the case $M_{\ell^*} = \Lambda$ the existence of excited electrons (muons) is excluded up to masses of 2.45 (2.47)\TeV at 95\% confidence level. Neutral current decays of excited leptons are considered for the first time, and limits are extended to include the possibility that the weight factors $f$ and $f^{\prime}$, which determine the couplings between standard model leptons and excited leptons via gauge mediated interactions, have opposite sign.
}

\hypersetup{%
pdfauthor={CMS Collaboration},%
pdftitle={Search for excited leptons in proton proton collisions at sqrt(s) = 8 TeV},%
pdfsubject={CMS},%
pdfkeywords={CMS, physics, excited leptons}}
\maketitle

\section{Introduction}
\label{sec:intro}

The standard model (SM) of particle physics describes the observed phenomena very successfully, however it provides no explanation for the three generations
of the fermion families.
Attempts to explain the observed hierarchy have
led to a class of models postulating that quarks and leptons may be composite objects of fundamental constituents \cite{Pati:1975md,compositeness1, Eichten1982, Eichten:1983hw, harari,
ssc-physics, Baur90,   Greenberg:1974qb, Greenberg:1980ri}.
The fundamental constituents are bound by an asymptotically free gauge interaction that becomes strong at a characteristic scale $\Lambda$. Compositeness models predict the existence of
excited states of quarks ($\PQq^{*}$) and leptons (\lstar) at the characteristic scale of the new binding interaction.
Since these excited fermions couple to the ordinary SM fermions, they could be produced via contact interactions (CI) in collider experiments,
with subsequent decay to ordinary fermions through
the emission of a $\PW/\PZ/\gamma$ boson, or via CI to other fermions.

Searches    at   LEP~\cite{Buskulic:1996tw,   Abreu:1998jw, Abbiendi:1999sa, Achard:2003hd},    HERA~\cite{H1estar}, and   the Tevatron~\cite{CDFestar,cdfmu,d0,D0estar} have found no evidence for excited
leptons. At  the Large  Hadron Collider (LHC) at CERN, previous searches  performed  by the CMS~\cite{cms-limit_new} and the ATLAS collaborations~\cite{atlas-limit_new} have also found no evidence of
excited  leptons,
obtaining a lower limit on the mass $M_{\ell^{*}}<2.2\TeV$ for the case $M_{\lstar}=\Lambda$.

In this paper, a search for excited leptons (\estar and \mustar) is presented, using a data sample of pp collisions at a center-of-mass energy $\sqrt{s} = 8\TeV$ collected with the CMS detector at the LHC in
2012 and corresponding to an integrated luminosity of $19.7\pm 0.5\fbinv$~\cite{CMS-PAS-LUM-13-001}. We consider the production of an excited lepton in association with an oppositely charged lepton of the
same flavor,
with subsequent radiative decays (\llg) or neutral current decays (\llZ).

\section{Theory and model assumptions}
\label{sec:theo}

The composite nature of quarks and leptons, if it exists, will manifest itself, above a characteristic energy scale $\Lambda$, as a spectrum of excited states. Such excited fermions, f$^*$, may couple to SM
leptons and quarks via a four-fermion CI that can be described by the effective Lagrangian

\begin{equation}
\mathcal{L}_{\mathrm{CI}} = \frac{g_{*}^{2}}{2\Lambda^{2}}j^{\mu}j_{\mu} ,
\end{equation}

where $\Lambda$ is the energy scale of the substructure, assumed to be equal to or larger than the excited fermion mass.
The quantities $g_{*}^{2}=4\pi$, and $j_{\mu}$, defined in Ref.~\cite{Baur90}, involve only left-handed
currents by convention. In addition to the coupling via CI, excited fermions can also interact with SM fermions via gauge interactions.
For excited leptons, the corresponding Lagrangian for the
gauge-mediated (GM) interaction is given by

\begin{equation}
\mathcal{L}_{\mathrm{GM}} = \frac{1}{2\Lambda}\bar{\mathrm{f}}_{R}^{*}\sigma^{\mu\nu} \left(gf \frac{\tau}{2} {W}_{\mu\nu} + g^{\prime}f^{\prime}\frac{Y}{2}B_{\mu\nu}
\right) \mathrm{f}_{L} + h.c.
\end{equation}

where ${W}_{\mu\nu}$ and $B_{\mu\nu}$ are the field-strength tensors of the SU(2) and U(1) gauge fields, and $g = e / \sin \theta_{W}$. The quantity, $g' = e / \cos \theta_{W}$
represents the electroweak gauge coupling with the Weinberg angle $\theta_{W}$, and $Y$ and $\tau$ are the generators of the U(1) and SU(2) groups, respectively.
The quantities $\mathrm{f}_R$ and $\mathrm{f}_L$ are the right and left-handed components of the lepton or excited lepton.
The weight factors $f$ and $f^{\prime}$
define the couplings between SM leptons and excited leptons via gauge interactions~\cite{Baur90}.
The compositeness scales contained in $\mathcal{L}_{\mathrm{CI}}$ and $\mathcal{L}_{\mathrm{GM}}$ are assumed to be the same.

The excited lepton, \lstar, can decay to a SM lepton via a CI $\ell^* \to \ell \mathrm{f\bar{f}}$, where $\mathrm{f}$ is a fermion, or through the mediation of a gauge boson via a gauge interaction.
The following gauge-interaction-mediated decays are possible: radiative decay $\lstar \to \ell \gamma$, charged-current decay $\lstar \to \ell^{\prime} \PW$, and
neutral-current decay $\lstar \to \ell  \PZ$. All four transitions, the CI and the three gauge interactions, are possible if $f = f^{\prime}$, while $f = -f^{\prime}$ forbids decays via photon emission. Since the exact relationship between the weight
factors is unknown, the results are interpreted for two extreme values: $f = f^{\prime} = 1$ and $f = -f^{\prime} = 1$.

\begin{figure}[thb]
\begin{center}
\includegraphics[width=0.49\textwidth]{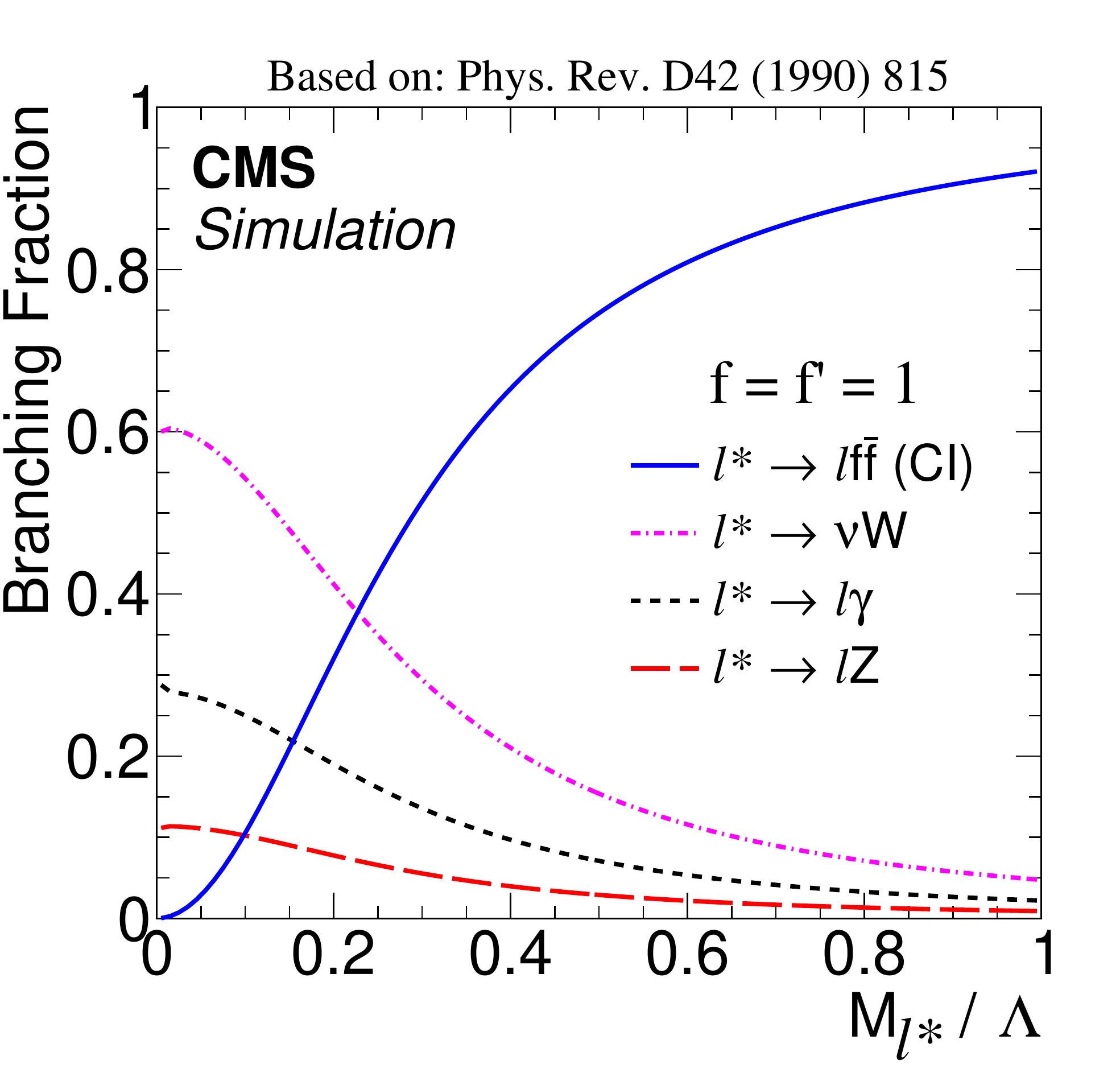}
\includegraphics[width=0.49\textwidth]{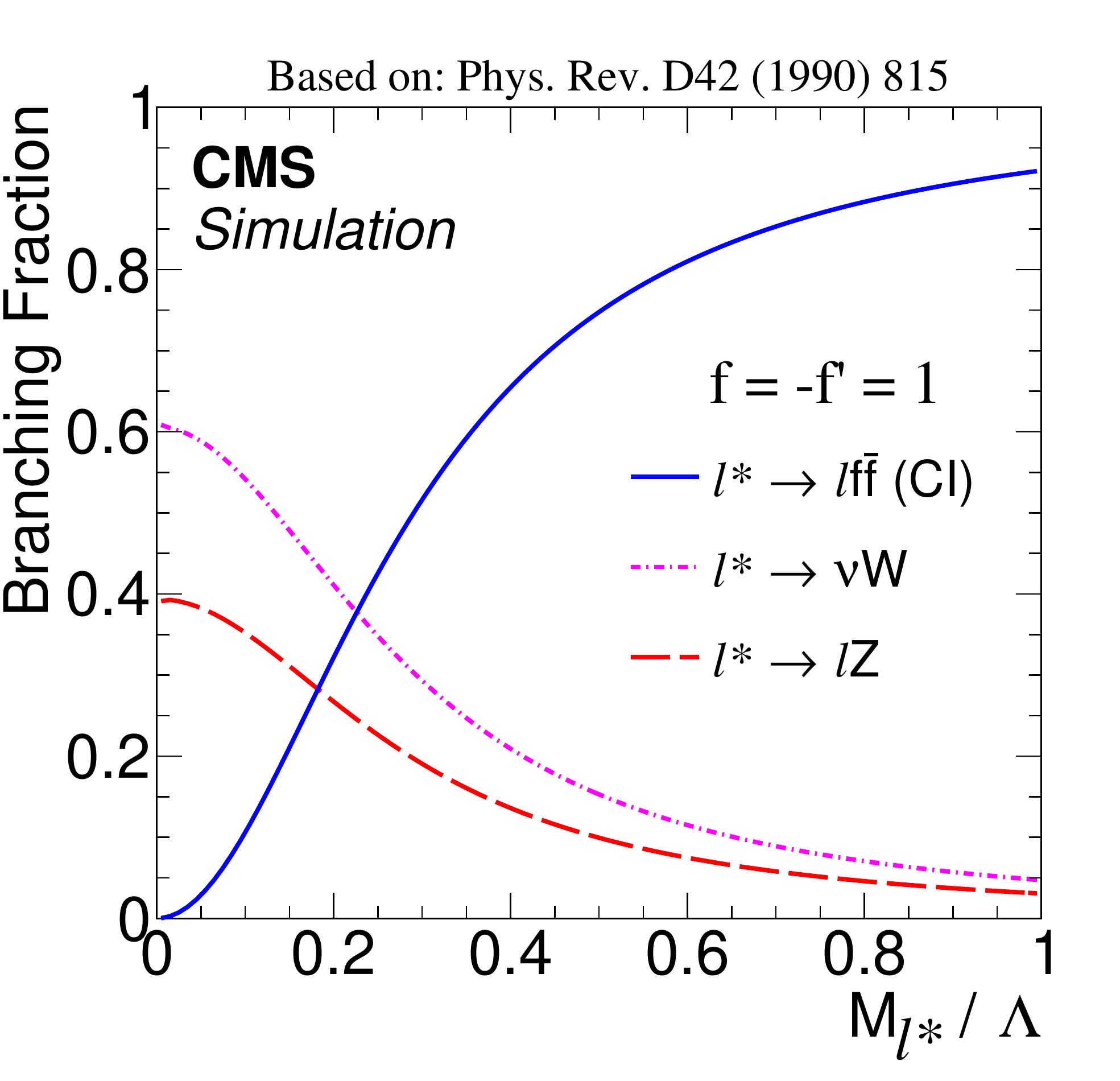}
\end{center}
\caption{Branching fractions for the decay of excited leptons, as a function of the ratio $M_{\ell^{*}} /
\Lambda$ of their mass to their compositeness scale, for the
coupling weight factors $f = f^{\prime} = 1$ (left) and $f = -f^{\prime} = 1$ (right).
The process $\lstar \to \ell\bar{f}f$ indicates the decay via CI, while the other
processes are gauge mediated decays.
}
\label{fig:AllChannels}
\end{figure}

In the present analysis we search
for the production of excited electrons and muons, \estar and \mustar, through a CI, which is dominant at the LHC for the model considered here. Excited leptons can also be
produced via
gauge
interactions, but those processes involve electroweak couplings and contribute less than $1\%$ to the cross section at the LHC; they have therefore been neglected here.
For light \lstar, the
decay of excited leptons via gauge interactions is dominant, while the decay via a CI becomes dominant at high masses, as shown in Fig.~\ref{fig:AllChannels}. The decay via a CI is not
considered in the simulated samples used here.

\begin{figure}[thb]
\begin{center}
\includegraphics[height=0.3\textwidth]{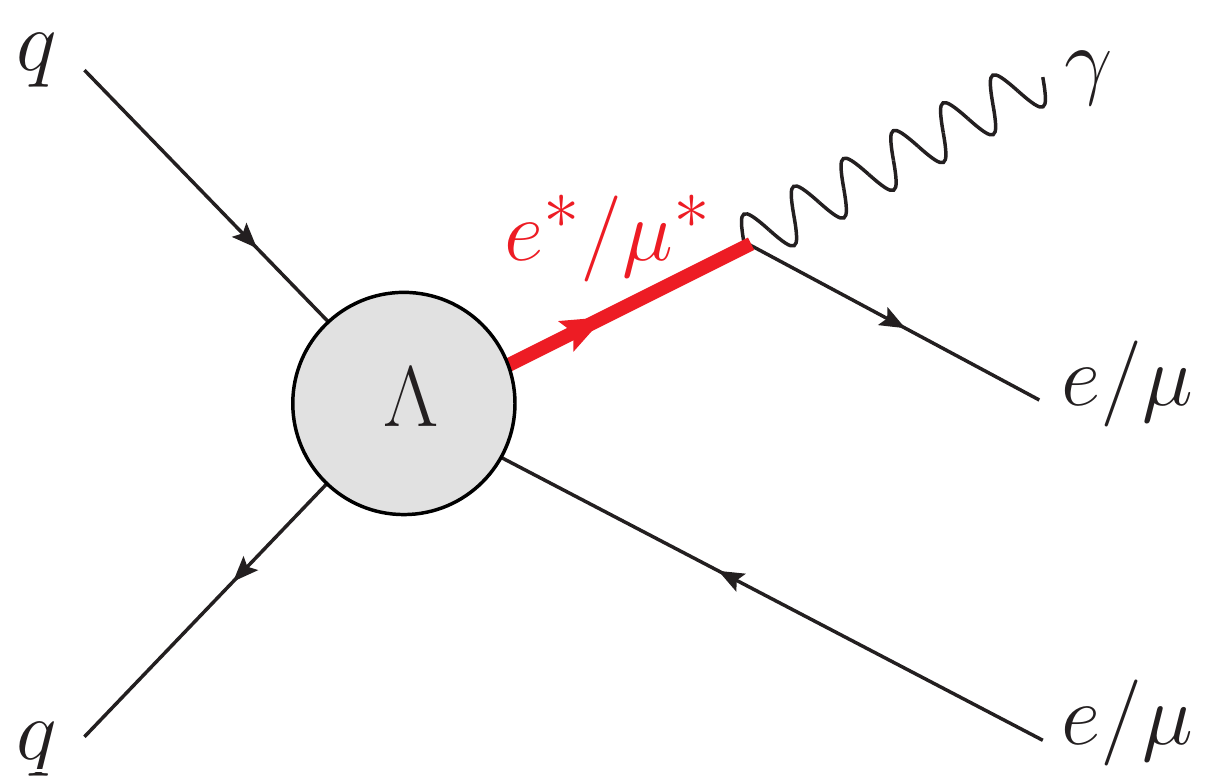}
\includegraphics[height=0.3\textwidth]{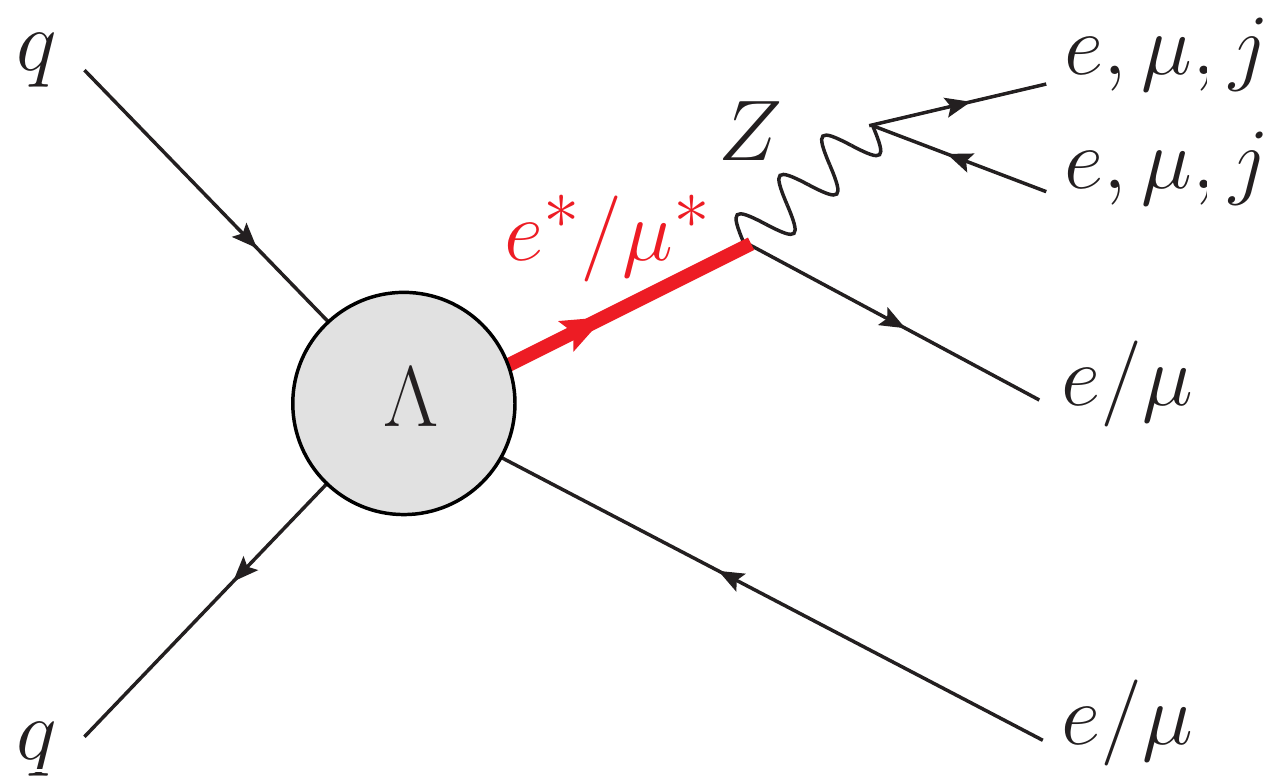}
\end{center}
\caption{Illustrative diagrams for \llg (left) and \llZ (right), where $\ell = \Pe$, $\Pgm$.
Decays of the $\PZ$ boson to a pair of electrons, muons or jets are considered.}
\label{fig:feynman}
\end{figure}

The search channels considered in this analysis are summarized in Table~\ref{tab:channels}. The \llg final state is represented by the Feynman diagram in Fig.~\ref{fig:feynman} left. A second class of
searches
seeks decays via the emission of a $\PZ$ boson (Fig.~\ref{fig:feynman} right), with the $\PZ$ boson decaying to either a pair of electrons, a pair of muons, or a pair of jets. This decay mode allows the
phase space where $f = -f^{\prime}$, unexplored by previous LHC searches, to be investigated. The transverse momentum (\pt) of the $\PZ$ boson coming from the decay of the excited lepton is larger for
heavier excited-lepton masses, and at high \pt the final-state particles are highly collimated. This characteristic is exploited in the \lljj decay mode, in which jet substructure techniques are used to
reconstruct a ``fat jet" corresponding to the $\PZ$ boson, and in the leptonic channels where the lepton isolation is modified.

Signal samples for both \estar and \mustar are produced using {\PYTHIA}8.153~\cite{Sjostrand:pythia8, Sjostrand:2006za}, which uses the leading order (LO) compositeness model described in Ref.~\cite{Baur90}.
Thirteen \lstar mass points from 200 to 2600\GeV have been simulated for all channels except the $\ell\ell \Pj\Pj$ channels, which starts at 600 GeV because of the analysis thresholds. Masses below 200 GeV
are
excluded
by previous searches at 95\% confidence level. All simulated events have been passed through the detailed simulation of the CMS detector based on {\GEANTfour}~\cite{geant4} and have been re-weighted so that
the distribution of
pileup events (contributions from additional pp interactions in the same bunch crossing) matches that measured in data. The signal cross sections are calculated with {\PYTHIA}8, and are corrected using the branching fraction to the
3-body decays
via CI as predicted in
Ref.~\cite{Baur90}, as this decay mode is not implemented in {\PYTHIA}. The factorization and renormalization scales are set to the mass square of the excited lepton ($M_{\ell^*}^{2}$), $\Lambda$ is set to 10 \TeV, and the  CTEQ6L1~\cite{cteq} parametrization for the parton distribution functions (PDF) is used. This particular choice of the value
of $\Lambda$ has no impact on the resulting kinematic distributions. Only the width of the \lstar resonance and the
\lstar
production cross section depend on $\Lambda$. As long as the width of the \lstar is small compared to the
mass
resolution of the detector, the signal efficiency is independent of $\Lambda$. Mass-dependent next-to-leading order (NLO) k-factors ranging from 1.2 to 1.35~\cite{kfactor}
are applied on the signal event yields. Production cross sections for the signals, as well as those of the different decay modes including the corresponding branching fractions are given in Table~\ref{tab:xsec}.

\begin{table}[t]
\renewcommand{\arraystretch}{1.1}
\centering
\topcaption{Final states for excited lepton searches considered in this analysis, where $\ell = \Pe$, $\Pgm$.
The notation for a specific channel is provided in the right most column. For neutral currents, the last two characters in this notation refer to particles from the decay of the $\PZ$ boson.
}
\label{tab:channels}
\begin{tabular}{c| l| c}
\hline
Decay mode      & \multicolumn{1}{c|}{Search channel} & Notation \\
\hline
\multirow{2}{*}{
\begin{tabular}{c}Radiative decay \\ $ \ell \ell^* \to \ell \ell \gamma $  \end{tabular}
}
& $ \Pe\Pe^* \, \, \to \Pe\Pe\gamma $ & $\Pe\Pe\gamma$  \\
& $ \Pgm\Pgm^* \to \Pgm\Pgm\gamma$ & $\Pgm\Pgm\gamma$ \\
\hline
\multirow{6}{*}{
\begin{tabular}{c} Neutral current \\ $ \ell \ell^* \to \ell \ell \PZ$ \end{tabular}
}
& $\Pe\Pe^* \, \, \to \Pe\Pe\PZ \, \to 4\Pe$        & $4 \Pe$        \\
& $\Pe\Pe^* \, \, \to  \Pe\Pe\PZ \, \to  2\Pe 2\Pgm$ & $2 \Pe 2 \Pgm$ \\
& $\Pe\Pe^* \, \, \to \Pe\Pe\PZ  \, \to  2\Pe 2\Pj $       & $2 \Pe 2 \Pj$   \\
& $\mu\mu^*  \to \Pgm\Pgm\PZ  \to 4\Pgm$         & $4 \Pgm$     \\
& $\Pgm\Pgm^* \to \Pgm\Pgm\PZ  \to 2\Pgm 2\Pe$          & $ 2 \Pgm 2 \Pe$ \\
& $\Pgm\Pgm^* \to \Pgm\Pgm\PZ  \to 2\Pgm 2\Pj$      & $2 \Pgm 2\Pj$ \\
\hline
\end{tabular}
\end{table}

\begin{table}
\renewcommand{\arraystretch}{1.1}
\centering
\topcaption{Excited lepton production cross section, and product of cross section and branching fraction for each of the three processes investigated,
as a function of the mass of the excited lepton. The values of the
k-factors are taken from Ref.~\cite{kfactor}. The case $f = -f^{\prime} = 1$ does not apply to the \llg channel. }
\label{tab:xsec}
{
 \begin{tabular}{c|ccc|c}
 \multicolumn{5}{c}{Production cross sections for excited leptons} \\
  \hline
  \multirow{2}{*}{$M_{\ell^{*}}$ (\GeVns)} & \multicolumn{3}{c|}{LO $\sigma$ (pb)} & \multirow{2}{*}{NLO k-factor} \\
  & $\Lambda = M_{\ell^{*}}$ & $\Lambda = 4\TeV$ & $\Lambda = 10\TeV$ & \\\hline
   200 & $1.3\times 10^{\, 5}$ & 0.84 & $2.2\times 10^{-2}$ & 1.30 \\
   1000 & 25.1 & $9.8\times 10^{-2}$ & $2.5\times 10^{-3}$ & 1.27 \\
   1800 & 0.28 & $1.1\times 10^{-2}$ & $2.9\times 10^{-4}$ & 1.28 \\
   2600 & $6.3\times 10^{-3}$ & $1.1\times 10^{-3}$ & $2.9\times 10^{-5}$ & 1.35 \\
   \hline
 \end{tabular}
 \vspace{0.5cm}

 \begin{tabular}{c|ccc|ccc}
 \multicolumn{7}{c}{$\sigma_\mathrm{NLO} \, \mathcal{B}$ (\llg) (pb)} \\
  \hline
  \multirow{2}{*}{$M_{\ell^{*}}$ (GeV)} & \multicolumn{3}{c|}{$f = f^{\prime} = 1$} & \multicolumn{3}{c}{$f = -f^{\prime} = 1$} \\
  & $\Lambda = M_{\ell^{*}}$ & $\Lambda = 4\TeV$ & $\Lambda = 10\TeV$ & $\Lambda = M_{\ell^{*}}$ & $\Lambda = 4\TeV$ & $\Lambda = 10\TeV$ \\
  \hline
   200 & $3.9\times 10^{\, 3}$ & 0.36 & $9.4\times 10^{-3}$ &\NA&\NA&\NA\\
   1000 & 0.70 & $2.0\times 10^{-2}$ & $8.0 \times 10^{-4}$ &\NA&\NA&\NA\\
   1800 & $7.7\times 10^{-3}$ & $1.2\times 10^{-3}$ & $7.5\times 10^{-5}$ &\NA&\NA&\NA\\
   2600 & $1.9\times 10^{-4}$ & $7.1\times 10^{-5}$ & $6.0\times 10^{-6}$ &\NA&\NA&\NA\\
  \hline
 \end{tabular}
 \vspace{0.5cm}

 \begin{tabular}{c|ccc|ccc}
 \multicolumn{7}{c}{$\sigma_\mathrm{NLO} \, \mathcal{B}$ (\lljj) (pb)} \\
  \hline
  \multirow{2}{*}{$M_{\ell^{*}}$ (GeV)} & \multicolumn{3}{c|}{$f = f^{\prime} = 1$} & \multicolumn{3}{c}{$f = -f^{\prime} = 1$} \\
   & $\Lambda = M_{\ell^{*}}$ & $\Lambda = 4\TeV$ & $\Lambda = 10\TeV$ & $\Lambda = M_{\ell^{*}}$ & $\Lambda = 4\TeV$ & $\Lambda = 10\TeV$ \\
   \hline
   200 & 772 & $7.2\times 10^{-2}$ & $1.9\times 10^{-3}$ & $2.7\times 10^{\, 3}$ & 0.28 & $7.3\times 10^{-3}$ \\
   1000 & 0.20 & $5.7\times 10^{-3}$ & $2.3\times 10^{-4}$ & 0.68 & $2.0\times 10^{-2}$ & $7.8\times 10^{-4}$ \\
   1800 & $2.2\times 10^{-3}$ & $6.8\times 10^{-4}$ & $2.1\times 10^{-5}$ & $7.5\times 10^{-3}$ & $1.2\times 10^{-3}$ & $7.4\times 10^{-5}$ \\
   2600 & $5.3\times 10^{-5}$ & $2.0\times 10^{-5}$ & $1.7\times 10^{-6}$ & $1.8\times 10^{-4}$ & $7.0\times 10^{-5}$ & $5.9\times 10^{-6}$ \\
  \hline
 \end{tabular}
 \vspace{0.5cm}

 \begin{tabular}{c|ccc|ccc}
 \multicolumn{7}{c}{$\sigma_\mathrm{NLO} \, \mathcal{B} (\ell\ell^{*} \to \ell\ell \PZ \to 2\ell 2\ell'$ ($\ell' = \Pe$, $\Pgm$)) (pb)} \\
  \hline
   \multirow{2}{*}{$M_{\ell^{*}}$ (\GeVns)} & \multicolumn{3}{c|}{$f = f^{\prime} = 1$} & \multicolumn{3}{c}{$f = -f^{\prime} = 1$} \\
    & $\Lambda = M_{\ell^{*}}$ & $\Lambda = 4\TeV$ & $\Lambda = 10\TeV$ & $\Lambda = M_{\ell^{*}}$ & $\Lambda = 4\TeV$ & $\Lambda = 10\TeV$ \\
    \hline
   200 & 73.5 & $6.8\times 10^{-3}$ & $1.8\times 10^{-4}$ & 256 & $2.6\times 10^{-2}$ & $6.9\times 10^{-4}$ \\
   1000 & $1.9\times 10^{-2}$ & $5.4\times 10^{-4}$ & $2.1\times 10^{-5}$ & $6.5\times 10^{-2}$ & $1.8\times 10^{-3}$ & $7.4\times 10^{-5}$ \\
   1800 & $2.1\times 10^{-4}$ & $3.4\times 10^{-5}$ & $2.0\times 10^{-6}$ & $7.2\times 10^{-4}$ & $1.1\times 10^{-4}$ & $7.0\times 10^{-6}$ \\
   2600 & $5.0\times 10^{-6}$ & $1.9\times 10^{-6}$ & $1.6\times 10^{-7}$ & $1.7\times 10^{-5}$ & $6.6\times 10^{-6}$ & $5.7\times 10^{-7}$ \\
  \hline
 \end{tabular}
 }
\end{table}

\clearpage

\section{The CMS detector}
\label{sec:detector}

The central feature of the CMS apparatus is a superconducting solenoid of 6\unit{m} internal diameter, providing a magnetic field of 3.8\unit{T}. Within the superconducting solenoid volume are a silicon
pixel and strip tracker, a lead tungstate crystal electromagnetic calorimeter (ECAL), and a brass and scintillator hadron calorimeter (HCAL), each composed of a barrel and two endcap sections. Forward
calorimeters extend the pseudorapidity~($\eta$)~\cite{JINST} coverage provided by the barrel and endcap detectors. Muons are measured in gas-ionization detectors embedded in the steel flux-return yoke
outside the solenoid.
In the barrel section of the ECAL, an energy resolution of about 1\% is achieved for unconverted or late-converting photons in the tens of GeV energy range. The remaining barrel photons have a resolution of
about 1.3\% up to $|\eta|=1$, rising to about 2.5\% at $|\eta|=1.4$. In the endcaps, the resolution of unconverted or late-converting photons is about 2.5\%, while the remaining endcap
photons have a resolution between 3 and 4\%~\cite{PhoRec}.
When combining information from the entire detector, the jet energy resolution amounts typically to 15\% at 10\GeV, 8\% at 100\GeV, and 4\% at 1\TeV, to be compared to about 40\%, 12\%, and 5\% obtained when the ECAL and HCAL calorimeters alone are used.
The electron momentum is determined by combining the energy measurement in the ECAL with the momentum measurement in the tracker. The momentum resolution for electrons with $\pt \approx 45\GeV$ from $\PZ
\to \Pe \Pe$ decays ranges from 1.7\% for non-showering electrons in the barrel region to 4.5\% for showering electrons in the endcaps~\cite{Khachatryan:2015hwa}.
Muons are identified in the range $\abs{\eta}< 2.4$, with detection planes made using three technologies: drift tubes, cathode strip chambers, and resistive plate chambers. Matching muons to
tracks measured in the silicon tracker results in a relative \pt resolution for muons with $20 <\pt < 100\GeV$ of 1.3--2.0\% in the barrel and better than 6\% in the endcaps. The \pt
resolution in the barrel is better than 10\% for muons with \pt up to 1\TeV~\cite{Chatrchyan:2012xi}.
A more detailed description of the CMS detector, together with a definition of the coordinate system used
and the relevant kinematic variables, can be found in Ref.~\cite{JINST}.

\section{Event selections}
\label{sec:objects}

\subsection{Triggers}

The selected trigger for each channel is summarized in Table~\ref{tab:object}.
For all channels, except those with a $2\Pe 2\Pgm$ final state, dilepton triggers are exploited:
the double electron trigger is used for events with electrons in the final state,
while muon events are selected by the dimuon trigger. Both triggers have identical \pt thresholds,
of 17 (8)\GeV for the leading (subleading) lepton.

\begin{table}[thbp]
\renewcommand{\arraystretch}{1.25}
\topcaption{Trigger requirement, offline $\pt$ and $\eta$-selection criteria, and event signature for all final state channels of the \lstar production and decay.}
\label{tab:object}
\begin{center}
\begin{tabular}{p{1.2cm}|p{2cm}|p{2.5cm}|p{3cm}|p{5.2cm}}
\hline
\multicolumn{1}{c|}{Channel} & \multicolumn{1}{c|}{Trigger} & \multicolumn{1}{c|}{Offline $\pt$} & \multicolumn{1}{c|}{Offline $|\eta|$} & \multicolumn{1}{c}{Signature and object ID} \\
\hline
$\newline \phantom{...} \Pe\Pe\gamma$ & Dielectron with 17(8)\GeV & $E^{\Pe 1}_{\mathrm{T}}> 35\GeV$, \newline $E^{\Pe 2}_{\mathrm{T}}> 35\GeV$, \newline $E^{\gamma \ }_{\mathrm T}> 35\GeV$ & $|\eta_{\Pe}| \ <1.44$, \newline $1.56 <
|\eta_{\Pe}| < 2.5$, \newline
$|\eta_{\gamma}| < 1.44$ & Two isolated high $E_{\mathrm T}$ electrons and one isolated high $E_{\mathrm T}$ photon \\
\hline
$\newline \phantom{...}  \Pgm\Pgm\gamma$ & Dimuon with 17(8)\GeV & $p^{\Pgm1}_{\mathrm T}> 35\GeV$, \newline $p^{\Pgm2}_{\mathrm T}> 35\GeV$, \newline $E^{\gamma \ }_{\mathrm T}> 35\GeV$ & $|\eta_{\Pgm}|<2.1$, \newline $|\eta_{\gamma}| < 1.44$ & Two
isolated high $\pt$ muons and one isolated high $E_{\mathrm T}$ photon \\
\hline
$\newline \phantom{...}  2\Pe2 \Pj$ & Dielectron with 17(8)\GeV & $E^{\Pe1}_{\mathrm T}> 35\GeV$, \newline $E^{\Pe2}_{\mathrm T}> 35\GeV$, \newline $E^{\Pj }_{\mathrm T} \ > 200\GeV$ &
$|\eta_{\Pe}| \, <1.44$, \newline $1.56 <
|\eta_{\Pe}| < 2.5$, \newline
$|\eta_{\Pj}|  \  < 2.4$ & Two isolated high $E_{\mathrm T}$ electrons and two jets that are merged from boosted $\PZ$ boson decays \\
\hline
$\newline \phantom{...} 2 \Pgm 2\Pj$ & Dimuon with 17(8)\GeV & $p^{\Pgm1}_{\mathrm T}> 35\GeV$, \newline $p^{\Pgm1}_{\mathrm T}> 35\GeV$, \newline $E^{\Pj }_{\mathrm T} \ > 200\GeV$ &
$|\eta_{\Pgm}|<2.4$, \newline $|\eta_{\Pj}| \  < 2.4$ & Two
isolated high
$\pt$ muons and two jets that are merged from boosted $\PZ$ boson decays \\
\hline
$\newline \newline \phantom{...} 4\Pe$ & Dielectron with 17(8)\GeV & $E^{\Pe \ }_{\mathrm T}> 25\GeV$ \newline for all four \newline electrons & $|\eta_{\Pe}|<1.44$, \newline $1.56 < |\eta_{\Pe}| <
2.5$ & Two isolated high $E_{\mathrm T}$ electrons
and two
nearby high $E_{\mathrm T}$ electrons from boosted $\PZ$ boson decay, using modified isolation for $\PZ$ boson decay electrons \\
\hline
$\newline \newline \phantom{...} 2\Pe 2\Pgm$ & Muon-Photon with 22\GeV each & $\pt>$ 25 GeV \newline for all four \newline leptons & $|\eta_{\Pe}| \, <1.44$, \newline $1.56 < |\eta_{\Pe}| < 2.5$, \newline
$|\eta_{\Pgm}| < 2.4$ & Two isolated
high $E_{\mathrm T}$ electrons and two nearby high $\pt$ muons from boosted $\PZ$ boson decay, using modified ID for one $\PZ$ boson decay muon and modified isolation for both Z
boson decay muons \\
\hline
$\newline \newline \phantom{...} 2\Pgm 2 \Pe$ & Muon-Photon with 22\GeV each & $\pt>$ 25 GeV \newline for all four \newline leptons & $|\eta_{\Pe}|<1.44$, \newline $1.56 < |\eta_{\Pe}| < 2.5$, \newline $|\eta_{\Pgm}| < 2.4$ & Two isolated high
$\pt$ muons and two nearby high $E_{\mathrm T}$ electrons from boosted $\PZ$ boson decay, using modified isolation for both $\PZ$ boson decay muons \\
\hline
$\newline \newline \phantom{...}  4\Pgm$ & Dimuon with 17(8)\GeV & $p^{\Pgm}_{\mathrm T}> 25\GeV$ \newline for all four \newline muons & $|\eta_{\Pgm}|<2.4$ & Two isolated high $\pt$ muons plus two nearby high $\pt$
muons from boosted Z
boson decay, using modified ID for one and modified isolation for both muons from $\PZ$ boson decay \\
\hline
\end{tabular}
\end{center}
\end{table}

The two cross channels with $2\Pe$ and $2\Pgm$ in the final state exploit a muon--photon trigger with a \pt threshold of 22\GeV for both objects,
where the photon trigger selects either electrons (as needed for this analysis) or photons, since the tracking information is not used at trigger level.
The muon--photon trigger is chosen because
the isolation requirements of the muon--electron trigger lead to an inefficiency when the two electrons from the $\PZ$ boson decay are close together. The trigger efficiencies are close to one in all cases
because of the large number of possible trigger objects. The offline \pt thresholds are set to 35\GeV for both electrons and muons, except for the 4-lepton channels, which require 25\GeV for each
lepton.

\subsection{Object reconstruction and selection}

\subsubsection{Electrons}
\label{sec:eles}

Electron candidates are identified as clusters of energy deposited in the ECAL, associated with tracks measured with the silicon tracker \cite{Khachatryan:2015hwa}.
The deposited energy should be predominantely in the electromagnetic calorimeter. Thus a lower limit is set on
the ratio H/E where H stands for the energy deposited in the hadronic calorimeter and E for that
in the electromagnetic calorimeter.
These candidates must be within the
barrel or endcap fiducial regions with $|\eta| < 1.44$ or $1.56 < |\eta| < 2.50$, respectively and have a $\pt > 35\GeV$ (25\GeV in the $4\ell$-searches). A set of identification
requirements that are optimized for electrons with high transverse momenta~\cite{Zprime}, based on the profile of the energy deposition in the ECAL and the matching between the track and the cluster,
are imposed to remove jets misidentified as electrons. The \pt sum of all other tracks (excluding the electron footprint) in a cone of $\Delta R =
\sqrt{\smash[b]{(\Delta\eta)^{2} + (\Delta\phi)^{2}}} < 0.3$ (where $\phi$ is the azimutal angle in radians) around the track of the electron candidate must be less than 5\GeV, a
selection denoted as "tracker isolation". In computing the tracker isolation for electrons, tracks have to originate from within a distance $|d_\mathrm{z}| < 0.2\unit{cm}$
from the primary vertex.
This requirement reduces the impact of pileup interactions vetoing candidate events.
The sum of the transverse energy (\ET) of calorimeter energy deposits in the same cone, referred to as "calorimeter isolation", must be less than $3\%$ of the candidate's transverse energy. The
calorimeter isolation energy is corrected for
pileup by the subtraction of the average energy per unit area of ($\eta, \phi$), computed for each event
using the
\textsc{FastJet} package~\cite{Cacciari:2011ma}.

For the two electrons from the $\PZ$ boson decay (in the \eeee and \mmee channels), the tracker isolation and calorimeter isolation for each electron are modified to remove the contribution of the other electron
\cite{ZZToqqll}.

\subsubsection{Muons}
\label{sec:muons}

The muon candidates have to pass identification (ID) criteria that are optimized for the reconstruction of muons with high transverse momenta~\cite{Zprime, Chatrchyan:2012xi}. In the ''global muon``
reconstruction, muons are reconstructed within $|\eta| < 2.4$ by combining tracks from the inner tracker and the outer muon system. The following requirements are imposed: at least one hit in
the pixel
tracker; hits in more than five tracker layers; and the detection of the muon in at least two muon stations. Since the stations are separated by thick layers of iron, the latter requirement
significantly reduces the probability of a hadron being misidentified as a muon. The relative uncertainty in the muon \pt measurement must not exceed $30\%$. In order to reduce the cosmic ray
muon background, the transverse impact parameter of the muon track with respect to the primary vertex of the event is required to be less than 0.2\unit{cm}.
The primary vertex is chosen as the one with the highest $\Sigma \pt^{2}$ of all charged tracks associated with that vertex. Furthermore, the muon is required to be isolated by demanding
that the scalar sum of the transverse momenta of all tracks, excluding the muon itself, within a cone of $\Delta R < 0.3$ around its own track, be less than $5\%$ of its \pt.

In the \eemm and \mmmm channels, one oppositely charged muon pair comes from the decay of the boosted $\PZ$ boson.
The muons can be close enough that one muon is inside the isolation cone of the other.
Therefore, for these
muons, the isolation calculation is modified by removing the contribution of the other muon.
Also, the identification requirements on one of these muons are loosened: the global muon
requirement is removed; the muon candidate is only required to be reconstructed in the tracker.
After these modifications, the reconstruction and identification efficiency of nearby muons are found to be comparable to those of separated muons \cite{ZZToqqll}. These two variations are referred to as
"modified identification" and "relaxed isolation".

\subsubsection{Photons}

For photons, identification criteria from Ref.~\cite{PhoRec} are applied to clusters in the ECAL that include requirements on the shower shapes, isolation
variables, and $H/E$ (ratio of deposits in
the HCAL and ECAL in a cone around the photon direction). A photon candidate is required to have a cluster with $\et> 35\GeV$ and to be in the barrel region of
the ECAL, with $\abs{\eta}<$ 1.44.
Photons are required to be in the central region because the jet-to-photon fake rate becomes high in the forward region, while only 4\% of a
signal would lie in this region. The
photon is also required to be isolated within a cone of radius $\Delta R < 0.3$ both in the tracker and the calorimeter. The cone axis is taken to be the
direction of the line joining the barycenter of
the ECAL clusters to the primary vertex. The isolation criteria depend on the $\eta$ of the photon, and distinguish between contributions from neutral and charged
hadrons and electromagnetic particles. As
with the electron isolation calculation, the sums do not include contributions from particles clearly associated with pileup vertices, and are adjusted for the
estimated residual pileup.

\subsubsection{Jets and \texorpdfstring{$\PZ \to \mathrm{jj}$}{Z to jj} tagging}
\label{sec:jets}
Hadronic jets are reconstructed from the list of particle flow (PF) candidates that
are obtained with the PF algorithm~\cite{CMS-PAS-PFT-10-001,CMS-PAS-PFT-10-001}, which reconstructs and identifies single
particles by combining information from all sub-detectors. Charged PF constituents not associated to the primary vertex are not used in
the jet clustering procedure. Good PF candidates are clustered into jets using the Cambridge-Aachen
(CA) algorithm~\cite{Wobisch:1998wt} with a distance parameter $R = 0.8$. An area-based correction is applied, to
take into account the extra energy clustered in jets from neutral particles in pileup
interactions, using the FASTJET software
package~\cite{Cacciari:2011ma}. Jet energy corrections are derived from the simulation, and are validated with in-situ
measurements using the energy balance of dijet, photon+jet, and $\PZ$+jets events~\cite{2011JInst...611002C}. Additional
quality criteria are applied to the jets in order to remove spurious jet-like features originating
from isolated noise patterns from the calorimeters or the tracker. These jet quality requirements are found to be 99\% efficient
for signal events. The jets are required to have $\pt>200\GeV$
and $|\eta| < 2.4$. Jets must also be separated from any
well-identified lepton (passing selections of Sections~\ref{sec:muons} and \ref{sec:eles}) by a cone of radius
$\Delta R > 0.8$.

In the $2\ell 2\Pj$ channels, the search is optimized for high-mass excited leptons that
produce a boosted, hadronically decaying $\PZ$ boson.
When such a highly boosted $\PZ$ decays to two quarks, their separation is often so small that they are reconstructed as a single jet with a mass larger than that of a typical Quantum ChromoDynamics (QCD) jet.
To achieve the best possible mass resolution for
this single jet, a {\it jet pruning} algorithm~\cite{jetpruning1,Ellis:2009me} is applied,
which is also used by the CMS collaboration for several other physics analyses
with hadronic decays of boosted $\PW$ and $\PZ$ bosons~\cite{CMS:EXO11006,CMS:SMP12019,CMS:EXO11095,ZZToqqll,CMS:JME13006}. This pruning procedure involves reclustering the constituents
of the original jet and applying additional requirements to eliminate
soft QCD radiation and large-angle QCD radiation coming from sources other than the $\PZ$ boson.
The kinematic distributions of the resultant jet are
a closer reflection of the hard process. In particular, the pruned jet
mass is closer to the mass of the parent $\PZ$ boson.

In addition, to further discriminate against jets from gluon and
single-quark hadronization, a quantity called N-subjettiness is used
~\cite{Thaler:2010tr,Thaler:2011gf,Stewart:2010tn}.
Before the pruning procedure is applied, the jet constituents are re-clustered with
the $k_{\mathrm{T}}$ algorithm~\cite{kt1,kt2}, until N joint
objects, called "subjets", remain in the iterative combination procedure of the
$k_{\mathrm{T}}$ algorithm. The N-subjettiness, $\tau_N$, is then defined as:

\begin{equation}
\tau_N = \frac{1}{\sum_{k} p_{\mathrm{T},k} R_{0}} \sum_{k} p_{\mathrm{T},k}\, \min( \Delta R_{1,k}, \Delta R_{2,k}...\Delta R_{N,k}),
\end{equation}

where the index $k$ runs over the jet constituents and the distances $\Delta R_{n,k}$ are calculated with respect to the axis of the $n^{th}$ subjet.
The quantity $R_{0}$ is set equal to the jet radius of the original jet.
The $\tau_N$ variable measures the capability of clustering the reconstructed particles in the jet
in exactly N-subjets: if it has a small value then it represents a
 configuration that is more compatible
with the N-subjettiness hypothesis.
In particular,
the variable that is best able to discriminate between the jets from a boosted
$\PZ$ boson decay and standard QCD jets
is the ratio of 2- to 1-subjettiness,
$\tau_{21}=\tau_{2} / \tau_{1}$.
If the jet has $\tau_{21} < 0.5$ and if its pruned mass falls in the range between 70--110\GeV, the jet is tagged as originating from a $\PZ$ boson, and is referred to as a "fat jet" in this paper.

The mismodeling of the $\tau_{21}$ variable can bias the signal efficiency estimated from the
simulated samples. A discrepancy between data and simulation has been observed
in the identification efficiency measured in events containing
merged jets produced by boosted $\PW$-bosons from top decays that pass
the same V-tag selections as the ones in this $\ell\ell \Pj \Pj$ analysis~\cite{ZZToqqll}.
Correction factors obtained from this sample are found to be $0.9\pm 0.1$.  These corrections are applied to
the signal efficiencies obtained from simulation.

\subsection{Signal selection}
\label{sec:selection}

In addition to the trigger and object identification requirements,
signal-candidate events are selected and SM backgrounds suppressed, sequentially as follows:

\begin{enumerate}
\item Selection of final state objects (see Section~\ref{sec:preselection}) and reconstruction of the boosted $\PZ$ boson in those channels containing a $\PZ$ boson.
\item Rejection of backgrounds with $\PZ$ bosons (see Section~\ref{sec:selection-Z}) with an invariant mass requirement.
\item Rejection of other backgrounds using a dedicated search window
(see Section~\ref{sec:selection-L}) that uses two calculations of \mlstar.
\end{enumerate}

\subsubsection{Preselection}
\label{sec:preselection}

As a first step, the final state objects are selected in the various search channels.

\begin{itemize}
\item {\llg:} Selection of two same flavor isolated leptons and one isolated high \ET photon within the acceptance and \pt thresholds given in Table~\ref{tab:object}. In the case of $\Pgm\Pgm\gamma$, muon pairs
that are back-to-back are rejected by removing those with an angle above $\pi -0.02$ to avoid contributions from cosmic ray muons. Additionally, the muons are required to have opposite charges. Selected
photons must be separated from the leptons by $\Delta R > 0.7$ to reduce the contribution from final state radiation.

\item {\lljj:} Selection of two isolated same flavor leptons and one fat jet (as defined in Section~\ref{sec:jets}) satisfying the acceptance and \pt thresholds given in Table~\ref{tab:object}. If
more than one fat jet is found, the one with the highest \pt is used. In the channel with muons, the muons are required to have opposite charges.

\item {\llll:} Selection of exactly four isolated leptons (four electrons, four muons or two electrons and two muons) within the acceptance and \pt thresholds given in Table~\ref{tab:object}.
First, the relaxed ID (for muons) and isolation are used for all leptons. Next, the boosted $\PZ$ boson is reconstructed. In the $2\Pgm 2\Pe$ ($2\Pe2\Pgm$) channel, the electron (muon) pair defines
the reconstructed $\PZ$ boson. In the $4\Pe$ and $4\Pgm$ channels, the lepton pair with invariant mass closest to the $\PZ$ pole mass is chosen. As a final step, the requirements on the leptons are
tightened. In channels with the boosted $\PZ$ boson decaying to muons, an additional charge requirement is applied  to both muons, and one of the muons originating from the $\PZ$ boson decay is allowed to
fulfill the relaxed ID only; all other leptons need to pass the full ID.

\end{itemize}

The invariant mass of the two leptons (in the $4\ell$ channels, of the two leptons that are not used to reconstruct the $\PZ$ boson) is denoted as $M_{\ell\ell}$ in what follows. This di-lepton mass is used to
reduce backgrounds that include $\PZ$ bosons not associated with the decay of putative heavy leptons.

\subsubsection{Invariant mass requirement}
\label{sec:selection-Z}

The invariant mass $M_{\ell\ell}$ is required to be above 106\GeV in the $\ell\ell\gamma$ and $4\ell$ channels, and above 200\GeV for the $2\ell 2\Pj$ channels, to reduce
backgrounds containing $\PZ$ bosons. This cut
efficiently removes contributions from $\PZ\gamma$ ($\PZ \PZ$) to the $\ell\ell\gamma$ and the $2\ell 2\Pj$ backgrounds.
For the $\Pe\Pe\gamma$ channel, there is an additional $\PZ$-veto on the two possible electron-photon
invariant masses to
remove electron pairs coming from a $\PZ$ decay, where one electron is misidentified as a photon.
Events are removed where any of the electron-photon invariant masses is within ${\pm}25\GeV$ of the nominal $\PZ$ boson mass.

\section{Modeling of the background}
\label{sec:samples}

\subsection{Sources of background}

Several SM processes contribute to the expected background for the various channels. Those contributions are discussed in the following.

\begin{itemize}

\item \llg channels:
Drell-Yan (DY) production is the most important background for the \llg channels, mostly originating
from production of a photon in association with a $\PZ$, which has a very similar signature to the signal.
It is simulated using
{\SHERPA}1.4~\cite{Gleisberg:2008ta} and its production cross section is normalized using a
NLO cross section calculated with the Monte Carlo (MC) program {\MCFM}6.1\&6.6~\cite{MCFM:2003, MCFM:2007}.

Subleading contributions to the background arise from diboson events with an additional high energy photon or events in which an electron is misidentified as a photon. Such
events are simulated using {\PYTHIA}6.4~\cite{Sjostrand:2006za}. Background contributions also arise from events
in which an additional prompt photon is produced together with a top pair ($\ttbar$+$\gamma$). These events are simulated with {\MADGRAPH}5.1~\cite{madgraph} using a LO cross section. All
these irreducible backgrounds arising from two prompt leptons and a prompt
photon are estimated using MC simulation. Smaller contributions due to events with two genuine leptons and a jet which has been misidentified as a photon are estimated
from data (see Section~\ref{sec:datadriven}). For the $\Pe\Pe\gamma$ channel, jets faking electrons may contribute, although at a negligible level (see Section~\ref{sec:datadriven-photons} for details). The
contribution of muons faked by jets is negligible.

\item \lljj channels:
The production of a $\PZ$ boson (decaying to leptons) plus additional jets is the dominant background followed by
the production of two top quarks and diboson events.
These contributions have been estimated from data, using simulation to validate the data-driven method described in
Section~\ref{sec:datadriven-jets}. All background contributions from simulation ($\ttbar$, diboson and DY+jets) are simulated using {\MADGRAPH} with NLO cross sections that were
calculated using {\MCFM}.

\item \llll channels:
The production of $\PZ\PZ$ (including $\PZ\gamma^{*}$), with both bosons decaying leptonically, is the main background to the four-lepton channel and contributes about $90\%$ of the total background expected.
An additional smaller contribution arises from the production of three vector bosons where some
of the leptons escape detection. The production of two top quarks, $\ttbar$, with or without an additional vector boson, can contribute to each channel. The background due to Higgs boson
production is negligible in the phase space considered here. In the four lepton channels, all the backgrounds have been estimated using predictions from simulations.
The $\PZ\PZ \to 4\ell$ background is described with \textsc{gg2zz}~\cite{gg2zz} for production via gluon fusion and
in the case of $\qqbar$ annihilation at NLO with
{\POWHEG}1.0~\cite{Nason:2004rx, Frixione:2007vw, Alioli:2010xd, Powheg:ZZ}.
Processes involving $\ttbar$+X (X = $\PW$, $\PZ$, $\gamma$) and triple boson samples are simulated with {\MADGRAPH}. It has been checked that the simulation describes correctly a sample of 4-lepton events selected as in Section \ref{sec:selection}, but relaxing the $\PZ$-vetoes to increase the number of events.

\end{itemize}

Table~\ref{tab:xsec_bkg} summarizes the simulated background samples with the corresponding NLO cross sections, and the channels where these samples are used.
{\PYTHIA} has been used to perform the fragmentation and hadronization of samples generated with {\MADGRAPH}.
The pileup simulation has been re-weighted so that the distribution of pileup events matches that measured in data.
All simulated events have been passed through the
detailed simulation of the CMS detector based on {\GEANTfour} \cite{geant4}.
Correction factors are also
applied to allow for differences between the simulated and measured reconstruction
efficiencies of the physics objects.

\begin{table}
\renewcommand{\arraystretch}{1.3}
\topcaption{Background samples with the corresponding generator and cross sections used for the various channels. Specific generator selections are shown where important
for the interpretation of the quoted cross sections.
}
 \begin{center}
\label{tab:xsec_bkg}
 \begin{tabular}{c|l|c|c|c}
 \hline
 Process & \multicolumn{1}{c|}{Selection} & Generator & NLO cross section (pb) & Channel \\
 \hline
 $\PZ$+jets $\to \ell\ell$+jets & $\pt^{\PZ} =$  70--100\GeV & MADGRAPH & $5.30 \times 10^{4}$ & $2\ell2$j \\
 $\PZ$+jets $\to \ell\ell$+jets & $\pt^{\PZ} > 100\GeV$   & MADGRAPH & $3.92 \times 10^{4}$ & $2\ell2$j \\
 $\PW$+jets $\to \ell\nu$+jets  &   \multicolumn{1}{c|}{\NA} & MADGRAPH & $3.63 \times 10^{4}$ & $2\ell2$j \\
 \hline
 Z$\gamma \to \ell\ell\gamma$ & $\Delta R(\gamma, \ell) > 0.6$ & SHERPA & 14.9 & $\ell\ell\gamma$ \\
 \hline
 $\ttbar$+jets &  \multicolumn{1}{c|}{\NA}  & MADGRAPH & 23.9 & $2\ell2$j \\
 $\ttbar \gamma$ & $E_{\mathrm T}(\gamma)>10\GeV$ &  MADGRAPH & 1.44(LO) & $\ell\ell \gamma$  \\
 $\ttbar$Z &   \multicolumn{1}{c|}{\NA}  & MADGRAPH & 0.208 & $4\ell$ \\
 $\ttbar$W &    \multicolumn{1}{c|}{\NA}  & MADGRAPH & 0.232 & $4\ell$ \\
 \hline
 WW $\to 2\ell 2\nu$ &    \multicolumn{1}{c|}{\NA}  & MADGRAPH & 6.03 & $2\ell2$j \\
 WW  &   \multicolumn{1}{c|}{\NA}  & PYTHIA6 & 54.8 & $\ell\ell \gamma$ \\
 \hline
 WZ $\to 2\ell2$q &  \multicolumn{1}{c|}{\NA} & MADGRAPH & 2.32 & $2\ell2$j, $4\ell$ \\
 WZ $\to 3\ell\nu$ & \multicolumn{1}{c|}{\NA}  & MADGRAPH & 1.00 & $2\ell2$j, $4\ell$ \\
 WZ & \multicolumn{1}{c|}{\NA} & PYTHIA6 & 33.2 & $\ell\ell \gamma$ \\
 \hline
 ZZ $\to 2\ell2$q & \multicolumn{1}{c|}{\NA}  & MADGRAPH & 2.47 & $2\ell2$j, $4\ell$ \\
 ZZ $\to 2\ell2\nu$ &  \multicolumn{1}{c|}{\NA}& MADGRAPH & 0.71 & $2\ell2$j \\
 ZZ $\to 4\ell$ & \multicolumn{1}{c|}{\NA} & MADGRAPH & 0.177 & $2\ell2$j \\
 ZZ inclusive & \multicolumn{1}{c|}{\NA} & PYTHIA6 & 17.7 & $\ell\ell \gamma$ \\
 ZZ $\to 4\ell$ & \multicolumn{1}{c|}{\NA} & POWHEG & 0.077 & $4\ell$ \\
 ZZ $\to 2\ell2\ell^{'}$ & \multicolumn{1}{c|}{\NA} & POWHEG & 0.176 & $4\ell$ \\
 gg $\to$ ZZ $\to 4\ell$ & \multicolumn{1}{c|}{\NA} & GG2ZZ & 0.005 & $4\ell$ \\
 gg $\to$ ZZ $\to 2\ell2\ell^{'}$ & \multicolumn{1}{c|}{\NA} & GG2ZZ & 0.012 & $4\ell$  \\
 \hline
 WWZ & \multicolumn{1}{c|}{\NA} & MADGRAPH & 0.063 & $4\ell$ \\
 WZZ &\multicolumn{1}{c|}{\NA}  & MADGRAPH & 0.020 & $4\ell$ \\
 ZZZ & \multicolumn{1}{c|}{\NA} & MADGRAPH & 0.005 & $4\ell$\\
 \hline
 \end{tabular}
 \end{center}
\end{table}

\subsection{Data-driven backgrounds}
\label{sec:datadriven}

\subsubsection{Misidentification of electrons}
\label{sec:datadriven-electrons}

 Backgrounds with zero or one real electron can contribute to the ee$\gamma$ candidate sample.
 The largest contributions come from processes such as
 $\PW (\to {\Pe}\nu)$+jet+$\gamma$ where the jet in the event is
 misidentified as an electron.
 Misidentification can occur when
 photons coming from $\pi^{0}$ or $\eta$ mesons inside a jet convert to an $\Pep\Pem$  pair.
 Other possible sources include processes with a charged particle within a jet providing a track in the tracker and an electromagnetic cluster that together fake an electron signature, or  a track
from a charged particle that matches a nearby energy deposition in the calorimeter from another particle.
 The misidentification rate, $f_\text{electron}^{\text{misid}}$,
 is calculated as the ratio between
 the number of candidates passing the electron selection criteria with respect to those satisfying looser selection criteria.
 The looser criteria require only that the first
 tracker layer contributes a hit to the electron track and that
 loose identification requirements on the shower shape
 and the ratio $H/E$ are satisfied.
The misidentification rate
is estimated as a function of
 \ET in bins of $\eta$ using a
 data sample selected with a trigger requiring at least one electromagnetic cluster.

 In order to estimate the contribution of misidentified electrons to
the selected events, the misidentification rate is applied to a
 subsample of data events containing one electron passing good electron
criteria and a second one passing a loose set of criteria. This loose set of criteria includes cuts on
shower shape  and the ratio $H/E$, but allows
one of the electron selection criteria to be missed.
The events are required to satisfy all other selection criteria of the analysis.

 The systematic uncertainty in $f_\text{electron}^{\text{misid}} (\ET, \eta)$
  is determined using a sample of events containing two reconstructed
 electrons as in~\cite{Zprime}.
 The  contribution from jet events to the inclusive dielectron mass spectrum
 can be determined
 either by applying the misidentification rate twice on events with two loose electrons
 or by applying the misidentification rate once on events with one fully identified electron and one
 loose electron. The first estimate lacks contributions from $\PW$+jets and $\gamma$+jets
  events while the second estimate is contaminated
 with DY events. These effects are corrected for using simulated samples.
 The observed difference of $30\%$ between the two estimates is taken
 as the systematic uncertainty in the jet-to-electron misidentification rate.

\subsubsection{Misidentification of photons}
\label{sec:datadriven-photons}

 Hadronic jets in which a $\pi^{0}$ or $\eta$ meson carries a significant fraction of the energy
 may be misidentified as isolated photons.
 Thus $\PZ$+jet(s) events are a potential background for the $\ell\ell \gamma$ search.
 The photon misidentification rate is measured directly from
 data using a data set collected using a single photon trigger.
To avoid trigger biases, the events must contain
at least one reconstructed super-cluster (energy deposit in the ECAL)
besides the one that fired the trigger.
In addition,
 the ratio of hadronic energy
 to the energy of that super-cluster is required to be less than $5\%$.
 The misidentification rate
 is defined as the ratio of the number of
 photon candidates that pass all the photon
 selection criteria (numerator) to the ones that pass a loose set of shower shape requirements
 but fail one of the photon isolation criteria (denominator).
 The numerator sample can have a contribution from
 isolated true photons.
The contamination is estimated using
 the distribution of
 energy-weighted shower width
 computed in units of crystal lateral dimension.
The shower shape for isolated true photons is obtained from a simulated sample.
 The shape of non-isolated photons is obtained from data by considering a background dominated region (side-band region) of the photon isolation variable.  The true photon fraction in the numerator is
estimated by
 fitting these two different shower shapes (signal and background templates) to the
 shower shape distribution of the numerator sample.
The photon misidentification rate is calculated in photon \ET bins. It decreases with increasing photon \ET and is at most of the order of a few percent.
As an example, for photons of \ET = 100\GeV the jets-to-photon misidentification rate is about 1.5\%.

 In order to estimate the contribution of misidentified photons to the selected events, the misidentification rate is applied to a subsample of data events that satisfy all selection criteria listed in Section~\ref{sec:selection} except that the photon candidate must pass a looser set of shower shape requirements and fail one of the photon isolation criteria.

There are two main sources of uncertainties in the determination of jet to photon misidentification rate. First, the shower shape of non-isolated photons is obtained from data in the side band region:
changing the side band region results in some change in the template for non-isolated photons. Second, the probability for a jet to fake a photon is different for quark and gluon jets and the fraction of
jets due to quarks may not be the same in the sample used to obtain the fake rate and in the sample where this fake rate is applied. Considering these two sources of uncertainties, a conservative systematic
uncertainty of $50\%$ is assigned, independently of the photon \et.

\subsubsection{Data-driven background in \texorpdfstring{$2\ell 2\Pj$}{2l2j}}
\label{sec:datadriven-jets}

\begin{figure}[h!]
\centering
\vspace{0.7cm}
\begin{tabular}{lr}
\includegraphics[width=0.4\textwidth]{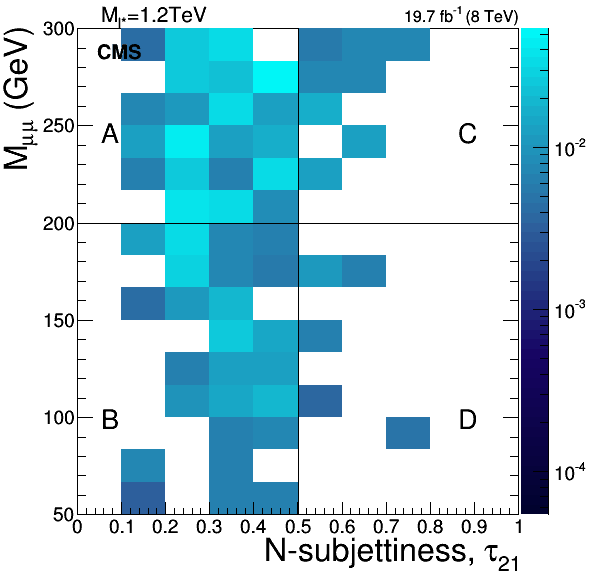} &
\includegraphics[width=0.4\textwidth]{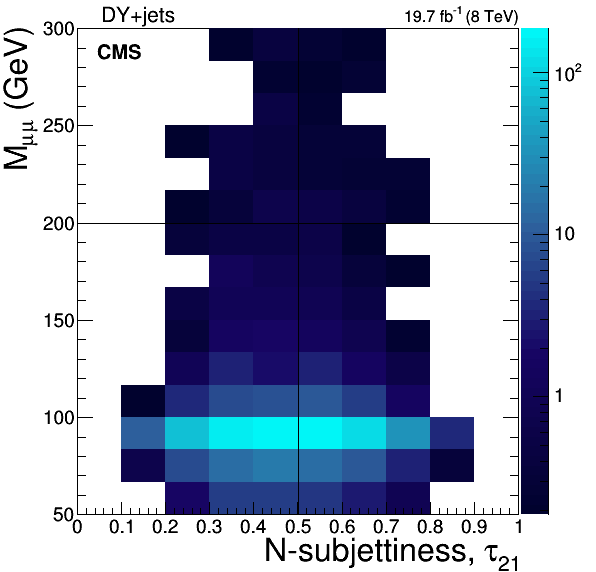} \\
\includegraphics[width=0.4\textwidth]{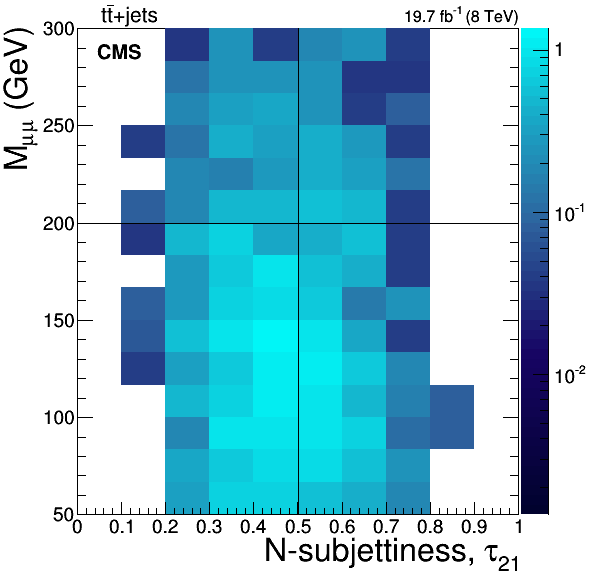} &
\includegraphics[width=0.4\textwidth]{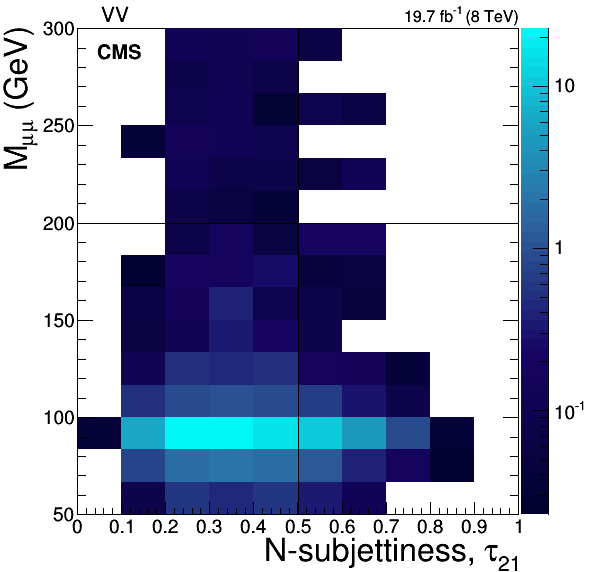} \\
\end{tabular}
\caption{Invariant mass of the pair of isolated muons vs. the N-subjettiness ratio, $\tau_{21}$, of the selected fat jet,
for events passing the selection criteria given in Section~\ref{sec:selection}, but with the cuts
on $M_{\ell\ell}$ and $\tau_{21}$ relaxed
for signal with $M_{\ell^*}=1.2\TeV$ (top left), Drell-Yan+jets (top right),
$\ttbar$+jets (bottom left) and diboson events (bottom right).
The right hand scale gives to the number of events corresponding to the given integrated
luminosity and the respective cross sections of the processes.
}
\label{abcdvar}
\end{figure}

The backgrounds due to DY+jets, $\ttbar$ and di-boson
 production are estimated using the ``ABCD'' method, which
relies on two variables to separate the
signal from the background. The two-dimensional
plane of these two variables is divided in four disjoint rectangular
regions A, B, C, and D, so that the region A is the signal
region, while B, C, and D are the control regions,
dominated by backgrounds.
If the ratio of the backgrounds
in regions A and B is the same as that for C and D (which holds if the two variables are independent), i.e.:
$N_{\mathrm{A}}/N_{\mathrm{B}}=N_{\mathrm{C}}/N_{\mathrm{D}}$, then
the background in the signal region A, $N_{\mathrm{A}}$ can be estimated
as:

\begin{center}
\begin{align}
N_{\mathrm{A}}=N_{\mathrm{C}}\frac{N_{\mathrm{B}}}{N_{\mathrm{D}}},
\label{eq:eqabcd}
\end{align}
\end{center}

where $N_{\mathrm{A}}$, $N_{\mathrm{B}}$, $N_{\mathrm{C}}$, and $N_{\mathrm{D}}$ are the background
events in regions A, B, C, and D, respectively.
The variables exploited in this analysis are the dilepton invariant mass $M_{\ell\ell}$
and N-subjettiness $\tau_{21}$.
The region A is defined by the selection cut given in Sections~\ref{sec:jets} and \ref{sec:selection}.
The regions B, C and D correspond to a similar selection
but with reversed requirements on $M_{\ell\ell}$ and/or
on the subjettiness ratio $\tau_{21}$ of the selected highest \pt fat jet.
These four regions are indicated in Fig.~\ref{abcdvar} (upper-left) along with
the borders of the regions (shown as solid lines) corresponding to the invariant mass
$M_{\ell\ell}$ being either above (for signal) or below (for background) 200\GeV and
$\tau_{21}$ being either above (background) or below (signal) 0.5.

For the $2\Pgm 2\Pj$ final state,
Fig.~\ref{abcdvar} shows background and signal predictions as a function of the invariant mass of the pair of isolated muons and the N-subjettiness ratio, $\tau_{21}$, of the fat jet but without selection on
$\tau_{21}$.
The background events displayed are from
simulated samples of DY+jets, $\ttbar$+jets, and diboson production,
while the signal events are from a sample simulated at $M_{\ell^*}=1.2\TeV$.
In the signal region A, about 20 background events are expected, with ${\sim}50\%$ originating from DY+jets and
${\sim}40\%$ due to $\ttbar$+jets.

Several tests were performed using simulated samples to verify that
the ABCD method using these variables reliably predicts the background yield.
The relation given in
Equation~\ref{eq:eqabcd} is expected to be independent of the choice of boundaries
for the control regions.
This assumption has been tested by applying the ABCD method to simulated
samples of DY+jets, tt+jets and di-boson events.
Moving the boundaries of regions B, C and D changed the calculated number of events in region A (whose definition is kept fixed) only slightly,
as shown in Table~\ref{changeboundry}.

\begin{table}[h]
\topcaption{\label{changeboundry} Events estimated in the region A
by applying the ABCD method to simulated samples of DY+jets, $\ttbar$+jets, and di-boson
events, as well as to data: each time with a different set of defining boundaries for regions B, C and D.
The true number of events in region A is $19.2 \pm 1.3$ and $15.0 \pm 1.1 $ for the muon
and electron channel, respectively.
}
\begin{center}
\begin{tabular}{l|c|c}
\hline
Thresholds & \multicolumn{1}{c|}{From simulation} & From data  \\
($M_{\ell\ell},~\tau_{21}$)  &  estimated $N_{\mathrm{A}}$ & estimated $N_{\mathrm{A}}$ \\ \hline
\multicolumn{3}{c}{muon channel}\\\hline
(200, 0.50)  & $19.3 \pm 1.4$ & $20.9 \pm 5.6$ \\
(180, 0.52) &  $19.8 \pm 1.6$ & $23.9 \pm 6.4$\\
(160, 0.54) &  $20.2 \pm 1.8$ & $20.4 \pm 6.4$\\\hline
  \multicolumn{3}{c}{electron channel}\\\hline
(200, 0.50) & $16.6 \pm 1.3$  & $ 22.1 \pm 5.9$  \\
(180, 0.52) & $16.3 \pm 1.3$  & $ 24.2 \pm 6.8$  \\
(160, 0.54) & $16.6 \pm 1.5$  & $ 23.9 \pm 7.2$  \\
\hline
\end{tabular}
\end{center}
\end{table}
\section{Systematic uncertainties}
\label{sec:systematics}

Three types of systematic uncertainties are considered:

\begin{itemize}
 \item {\bf Overall uncertainties in the simulation:} These include the uncertainty in the luminosity \cite{CMS-PAS-LUM-13-001}, the simulation of pileup and uncertainties in the cross sections used.
These uncertainties affect the normalization and are treated similarly for all background- and signal simulations.

 \item {\bf Object simulation uncertainties:} These depend on the final state of the respective analysis and are applied to the simulation of signal and background events. They consist, for example,
of uncertainties in the energy or momentum scale and resolution of the various particles, or in correction factors that were applied to account for differences between the simulated and the actual
performance of the detector.

 \item {\bf Uncertainties in background estimations from data:} These uncertainties are applied to background components that were estimated from data and are only relevant to the \llg and \lljj
channels.
\end{itemize}

The sources of these systematic uncertainties are discussed in Section~\ref{sec:uncertainties}
and their implications for signal and background in Section~\ref{sec:implications}.

\subsection{Object-specific simulation uncertainties}
\label{sec:uncertainties}

\subsubsection*{Electron uncertainties}
For electrons, uncertainties exist for the electron energy scale and electron identification efficiency. In both the barrel and the endcap, the scale uncertainties are determined by shifting the
transverse energy of the electrons by $0.3\%$~\cite{Khachatryan:2015hwa}. Systematic uncertainties due to electron identification are $2\%$ and $4\%$ \cite{Zprime} for very high energy electrons in barrel and endcap, respectively.

\subsubsection*{Muon uncertainties}
There are three sources of uncertainties for muons: uncertainties in the muon momentum scale; the muon momentum resolution; and the efficiency of the muon
selection. As
described in Ref.~\cite{Chatrchyan:2012xi} the uncertainty in the muon momentum scale is estimated to be $5\% \times \pt /\TeV$
and the effect of the scale uncertainty is estimated by changing the \pt by this value.
The uncertainty in the muon momentum resolution is 0.6\% and the effect of this uncertainty is estimated by smearing the \pt of the muons by an additional 0.6\%.
The uncertainty in the selection efficiency is $0.5\%$ for the identification criteria, and $0.2\%$ for the isolation criterion for each muon.

\subsubsection*{Photon uncertainties}
The energy scale and resolution uncertainties for photons are very small compared to those of the other objects. The
energy scale uncertainties are determined by shifting the transverse energies of the photons by $0.15\%$ in the barrel section of the calorimeter\cite{PhoRec}.

\subsubsection*{Jet-energy scale}
Jet-energy corrections are applied to account for the response function of the combined calorimeter system and other instrumental effects,
based on in situ measurements using dijet, $\PZ$+jet, and photon+jet data samples~\cite{2011JInst...611002C}.
Uncertainties due to these corrections are evaluated by shifting the jet energies by the calibration uncertainties (${\pm} 1\sigma$). The effect on signal yield was found to be less than $1\%$.

\subsection{Implications of uncertainties on signal and background yield}
\label{sec:implications}

The above sources of uncertainties are specifically associated with the simulation of the various objects.
To quantify each uncertainty on the signal and background, the relevant quantity is varied by ${\pm} 1 \sigma$, relative to the best estimate. Subsequently all kinematic selections are reapplied and the
impact on the analysis is
determined by calculating the difference of the result from that of  the original parametrization.

For all channels, the impact of pileup uncertainties was
estimated by shifting the mean number
of additional interactions and the inelastic cross section by $5\%$. The uncertainty in the signal yield cross section is taken to be $10\%$, following Ref.~\cite{kfactor}.

In the case of the four lepton final states, the dominant uncertainty in the background is the uncertainty in the cross section of the ZZ background, which is conservatively assumed to be $15\%$
(\cite{tagkey2015250}). Additional uncertainties with a large impact on the background yield are the electron energy scale (with impact on background yield of $12\%$), the electron selection efficiency ($6\%$), and the uncertainty in the electron resolution ($2.5\%$). The mixed channels suffer large effects from the electron energy scale ($8\%$), electron efficiencies ($5\%$), and muon efficiencies ($3\%$). In the four muon channel, the second largest uncertainty is associated with the muon selection efficiency ($4\%$) followed by that on the muon momentum scale ($1.6\%$). In this channel the uncertainties in the signal yield are completely dominated by the corresponding cross section uncertainty.

In the \llg channel, the dominant systematic uncertainty in the background is the uncertainty in the production cross section arising from the parametrization of the parton distribution functions in
the main background ($Z\gamma$),
which was determined to be $10\%$
by changing the choice of PDF set in the simulation
according to Ref.~\cite{Alekhin:2011sk,Botje:2011sn}.
Although the uncertainty in the data-derived background was determined to be $50\%$, its impact is rather small ($4\%$), as the total contribution of this background is rather small. The impact of the photon energy scale and resolution are negligible. One of the dominant systematic uncertainties for the signal in the \mmg channel is that in the muon momentum scale \cite{Chatrchyan:2012xi}, which rises with increasing \pt. As a result, the impact on the final event yield is rather small in case of the background, containing muons of moderate momenta,
but reaches more than $5\%$ in the high-mass signal samples.

In the \lljj channel, the dominant systematic uncertainty in the background is that associated with the background estimation method, mainly the signal contamination in control regions B, C and D of the ABCD matrix. This depends on the $M_{\ell^*}$ parameter; the lowest mass point represents the worst-case scenario where such contamination is maximal, and the highest mass point is the best-case scenario. The effect of the signal leakage in the control regions was estimated for various mass points between $M_{\ell^*} = $ 0.6 and 2.6\TeV, and found to be of the order of $30\%$ in the worst cases.
Another source of systematic uncertainties arises from the $\PZ$ tagging, since there is a discrepancy between the $\PZ$ tagging efficiency in data and simulation, as discussed in Section~\ref{sec:jets}. Based on the correction factors measured, a $10\%$ uncertainty is assigned the estimated signal
efficiency.

\newpage
\section{Final selection and results}
\label{sec:selection_final}

Table~\ref{tab:evtyield} summarizes the event yield for all channels after applying the selections for the leptons, photon or $\PZ$ boson, and the invariant mass cuts given in Section~\ref{sec:selection}.
Data agree with the SM expectation and no evidence for new physics is seen.

In the photon channels the main background ($\PZ\gamma$) contributes almost $90\%$ of the total. The remaining contributions are $\ttbar\gamma$ (${\lesssim} 7\%$) and jet/ photon misidentification (estimated from data to be ${\lesssim} 3\%$), and are rather small in comparison. Similarly in the four lepton channels, about $90\%$ of the backgound arises from $\PZ \PZ$. The jet channels have mixed composition. The main background
($\PZ$+Jets) contributes about $50\%$. The second largest contribution ($\ttbar$) contributes $40\%$ of the expected background.
The description of the background is based on the data driven approach described above, but the composition is estimated using simulation, since this information cannot be obtained from the data.

\begin{table}[htb]
{
\renewcommand{\arraystretch}{1.15}
 \topcaption{Expected background events, measured data events and expected signal yields for various channels before the L-shape optimization. Quoted uncertainties
are the quadratic sum of statistical and systematic errors. The signal yields are presented for different values of $\Lambda$, for the cases $f = f^{\prime} = 1$ and $f = -f^{\prime} = 1$. No signal is expected in
\llg for $f = -f^{\prime} = 1$.}
 \label{tab:evtyield}
 \begin{center}
 \begin{tabular}{c|c|c|c|c|c|c}
    \hline
    \multicolumn{1}{c|}{\multirow{4}{*}{Channel}} & \multirow{4}{*}{ $N_\mathrm{bg} \phantom{...}$} & \multirow{4}{*}{ $N_\mathrm{data} \phantom{..}$} & \multicolumn{2}{c|}{$N_\mathrm{signal}$}
     & \multicolumn{2}{c}{$N_\mathrm{signal}$}
    \\
     &  &  & \multicolumn{2}{c|}{$M_{\ell^{*}} = 0.6$ TeV} & \multicolumn{2}{c}{$M_{\ell^{*}} = 2$ TeV} \\
     &  &  & \multicolumn{2}{c|}{$f {=} f^{\prime} {=} 1 $ ($\, f {=} {-}f^{\prime} {=} 1\, $)} & \multicolumn{2}{c}{$f {=} f^{\prime} {=} 1 $ ($\, f {=} {-}f^{\prime} {=} 1\, $)} \\
     &  &  & $\Lambda = M_{\ell^{*}}$ & $\Lambda = 4\TeV$ & $\Lambda = M_{\ell^{*}}$ & $\Lambda = 4\TeV$ \\
    \hline
    \multirow{2}{*}{$\Pe\Pe\gamma$} & \multirow{2}{*}{$70.4 \pm 7.9$} & \multirow{2}{*}{62} & $1.1 \times 10^{\, 5}$ & $5.7 \times 10^{\, 2}$ & 25 & 5.1 \\
     &  &  & (0) & (0) & (0) & (0) \\
    \hline
    \multirow{2}{*}{2e2j} & \multirow{2}{*}{$22.1 \pm 6.0$} & \multirow{2}{*}{25} & $1.3 \times 10^{\, 4}$ & 69 & 4.7 & 1.0 \\
     &  &  & ($4.6\times 10^{\, 4}$) & ($2.4 \times 10^{\, 2}$) & (16) & (3.3) \\
    \hline
    \multirow{2}{*}{4e} & \multirow{2}{*}{$ \phantom{0} 3.0 \pm 0.6$} & \multirow{2}{*}{0} & $1.4 \times 10^{\, 3}$ & 7.5 & 0.3 & 0.1 \\
     &  &  & ($5.0 \times 10^{\, 3}$) & (26) & (1.1) & (0.2) \\
    \hline
    \multirow{2}{*}{2e2$\Pgm$} & \multirow{2}{*}{$ \phantom{0} 2.9 \pm 0.5$} & \multirow{2}{*}{4} & $1.8 \times 10^{\, 3}$ & 9.3 & 0.4 & 0.1 \\
     &  &  & ($6.2 \times 10^{\, 3}$) & (32) & (1.5) & (0.3) \\
    \hline
    \multirow{2}{*}{$\Pgm\Pgm\gamma$} & \multirow{2}{*}{$ 119 \pm 15$} & \multirow{2}{*}{150} & $1.2 \times 10^{\, 5}$ & $6.4 \times 10^{\, 2}$ & 26 & 5.4 \\
     &  &  & (0) & (0) & (0) & (0) \\
    \hline
    \multirow{2}{*}{$2\Pgm 2\Pj$} & \multirow{2}{*}{$20.9 \pm 5.6$} & \multirow{2}{*}{25} & $1.6 \times 10^{\, 4}$ & 85 & 5.9 & 1.2 \\
     &  &  & ($5.6\times 10^{\, 4}$) & ($2.9 \times 10^{\, 2}$) & (20) & (4.1) \\
    \hline
    \multirow{2}{*}{$2\Pgm 2\Pe$} & \multirow{2}{*}{$ \phantom{0} 2.5 \pm 0.4$} & \multirow{2}{*}{2} & $1.7 \times 10^{\, 3}$ & 9.0 & 0.4 & 0.1 \\
     &  &  & ($6.0 \times 10^{\, 3}$) & (31) & (1.3) & (0.3) \\
    \hline
    \multirow{2}{*}{$4 \Pgm$} & \multirow{2}{*}{$ \phantom{0} 4.0 \pm 0.6$} & \multirow{2}{*}{4} & $2.3 \times 10^{\, 3}$ & 12.1 & 0.5 & 0.1 \\
     &  &  & ($7.9 \times 10^{\, 3}$) & (42) & (1.8) & (0.4) \\
    \hline
 \end{tabular}
 \end{center}
}
\end{table}

\begin{figure}[tbp]
\begin{center}

 \includegraphics[width=0.48\textwidth]{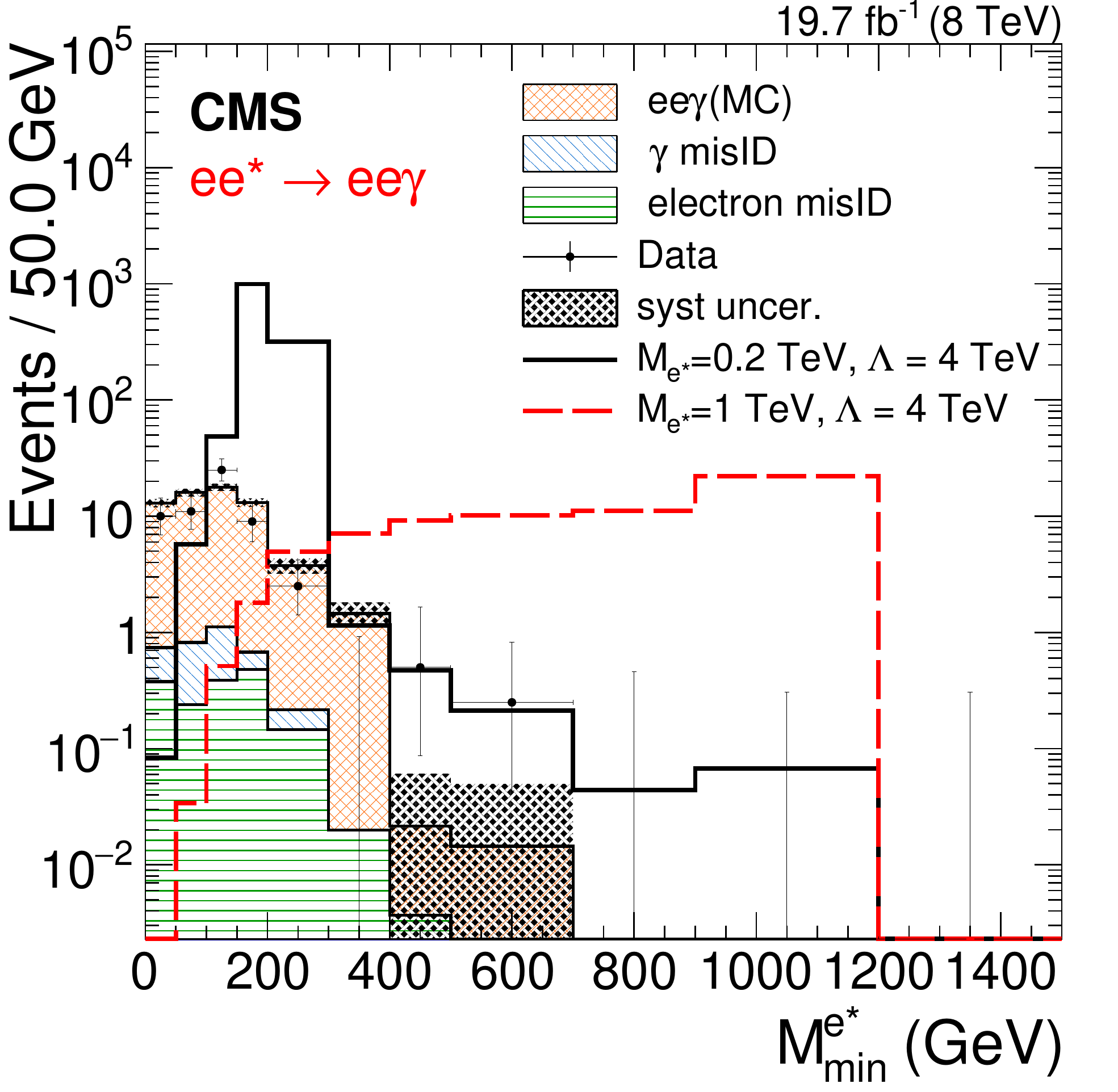}
 \includegraphics[width=0.49\textwidth]{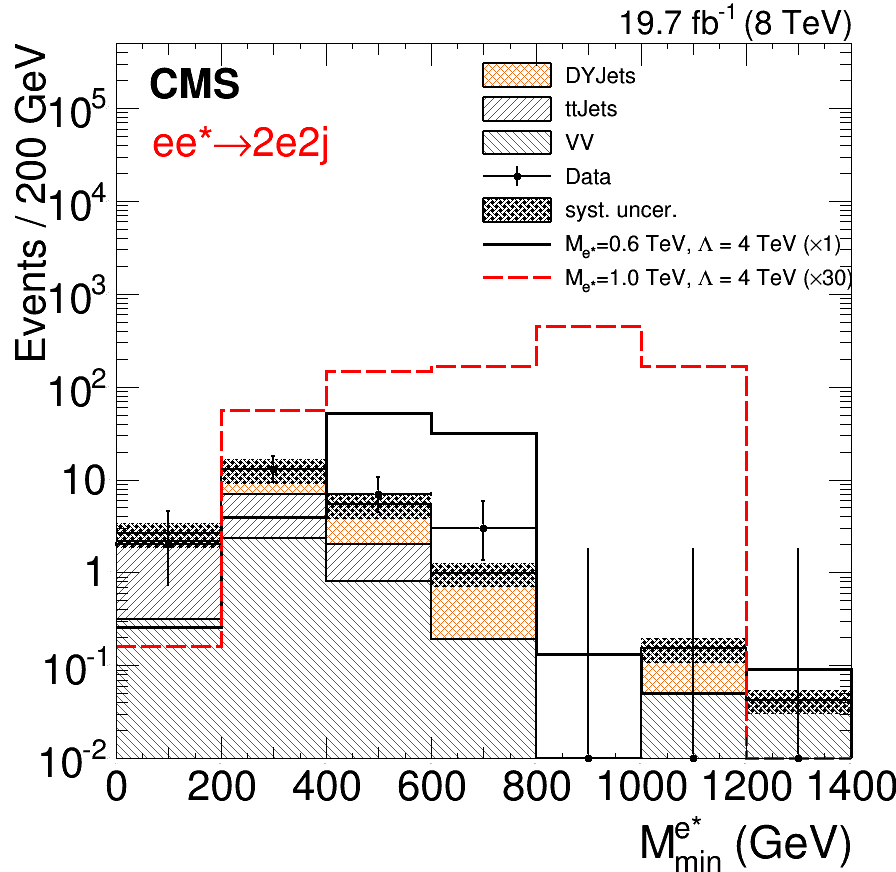} \\
 \vspace{1cm}
 \includegraphics[width=0.49\textwidth]{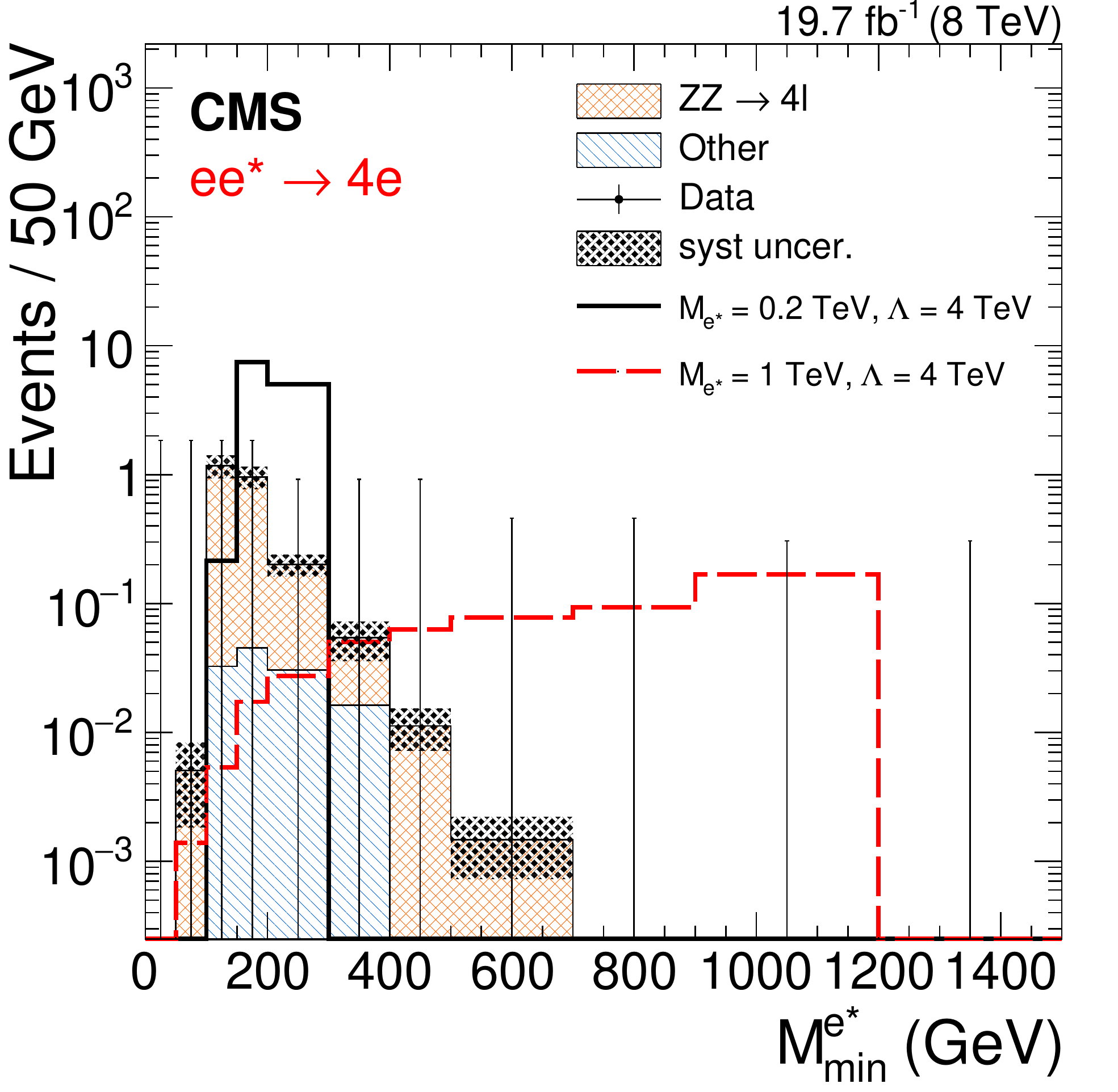}
 \includegraphics[width=0.49\textwidth]{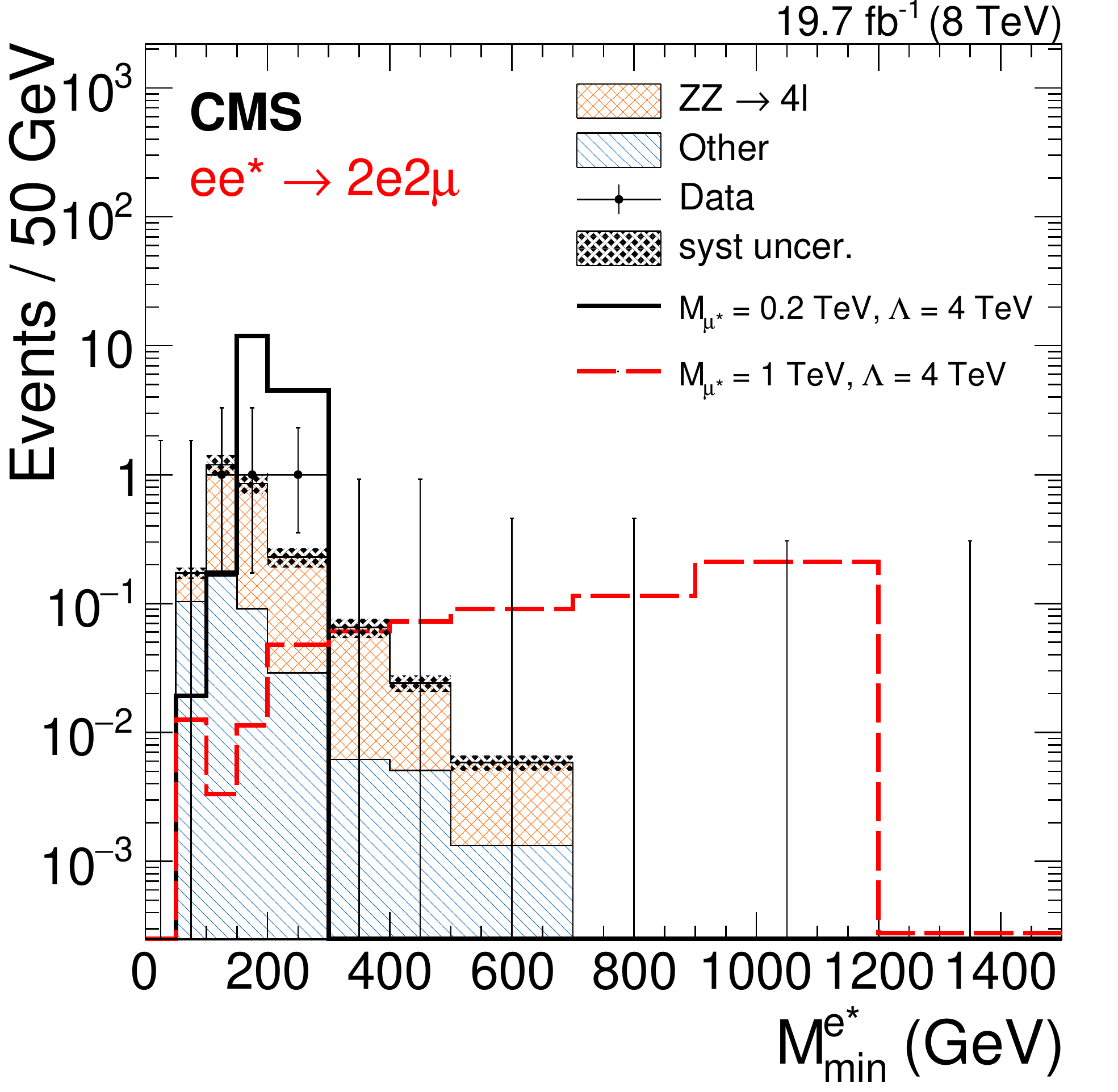}
 \end{center}
 \caption{Reconstructed minimum invariant mass from the vector boson ($\gamma$, $\PZ$) plus one electron for the four
excited electron channels. Top left: $\Pe\Pe\gamma$, top right: $2\Pe 2\Pj$, bottom left: $4\Pe$, bottom right: $2\Pe 2\Pgm$.
Two signal distributions are shown for \mlstar = 0.2 and 1\TeV, except the $2\Pe 2\Pj$ channel where the trigger threshold
only allows searches for $\mlstar > 0.5\TeV$.
The asymmetric error bars
indicate the central confidence intervals for Poisson-distributed data and are obtained from the
Neyman construction as described in Ref.~\cite{Garwood1936}.}
 \label{fig:Mmin_electrons}
\end{figure}

\begin{figure}[tbp]
 \begin{center}
 \includegraphics[width=0.465\textwidth]{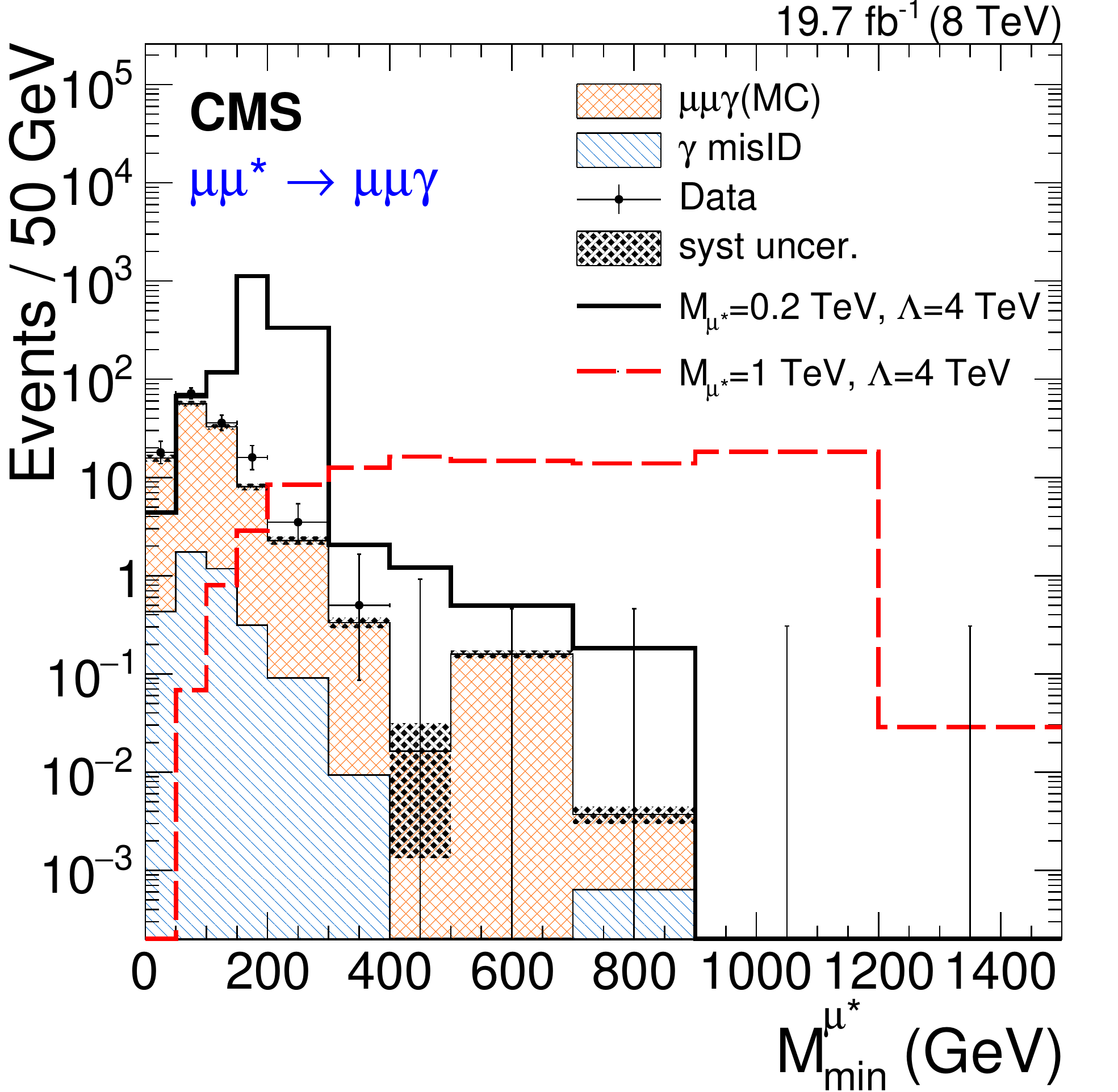}
 \includegraphics[width=0.50\textwidth]{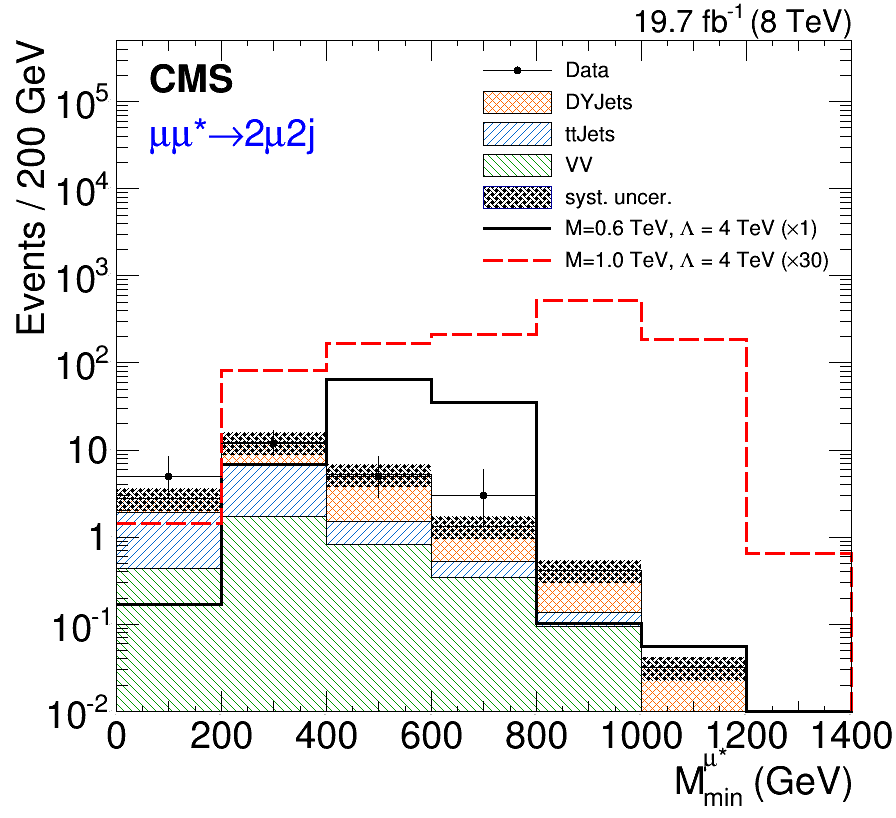} \\
 \vspace{1cm}
 \includegraphics[width=0.49\textwidth]{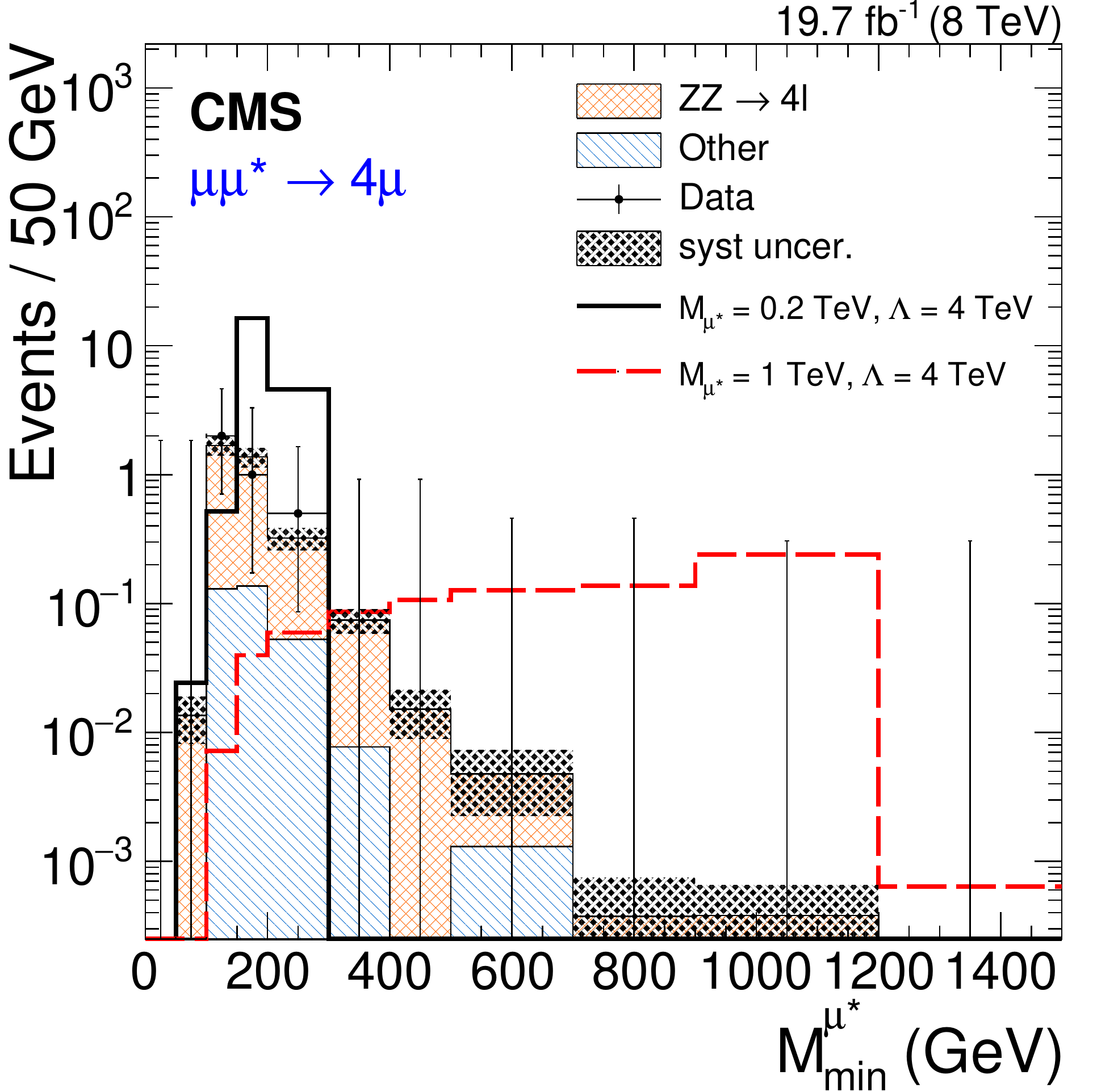}
 \includegraphics[width=0.49\textwidth]{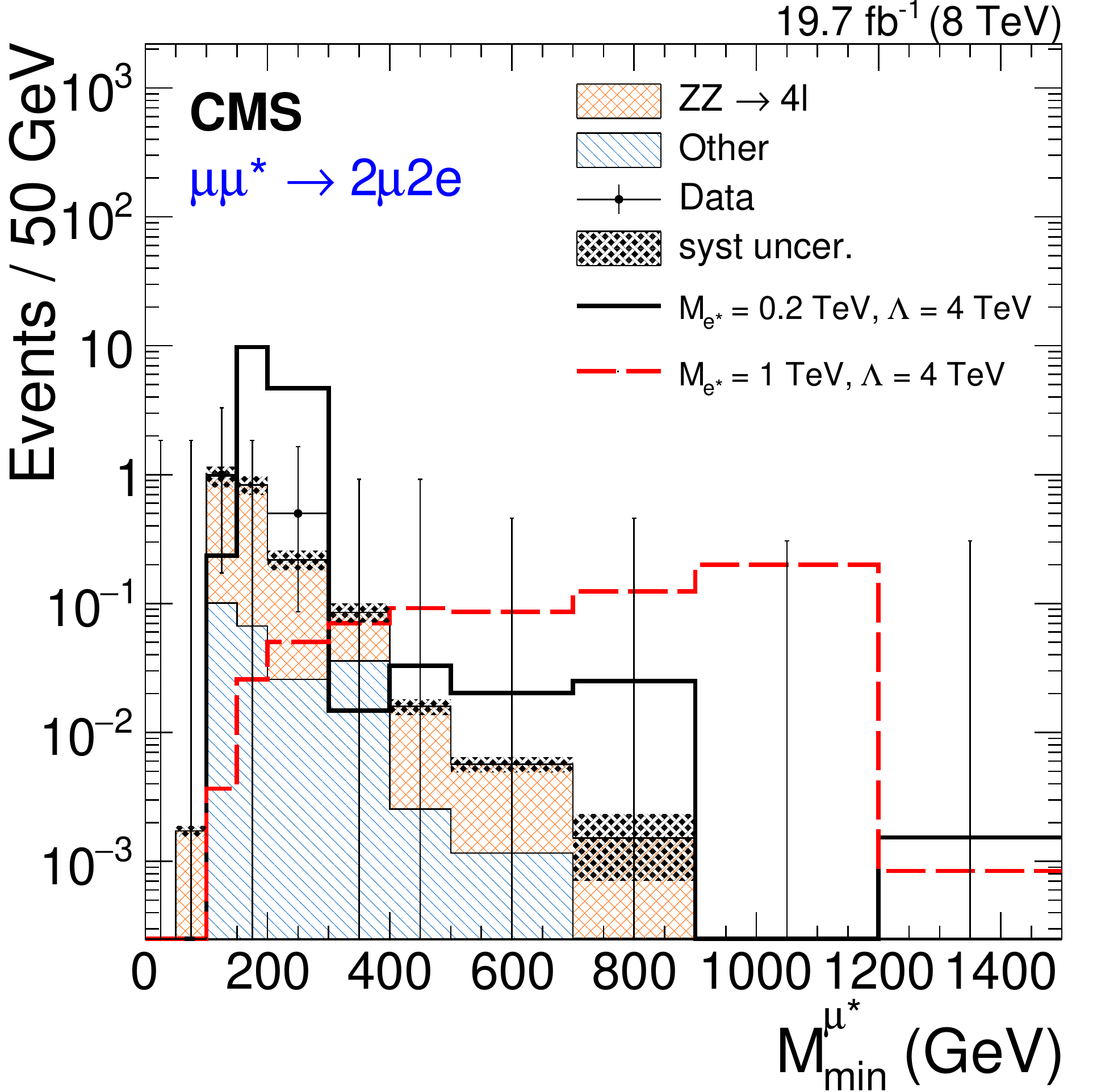}
 \end{center}
 \caption{Reconstructed minimum invariant mass from the vector boson ($\gamma$, $\PZ$) plus one muon for the four excited
muon channels. Top left: $\Pgm\Pgm\gamma$, top right: $2\Pgm 2\Pj$, bottom left: $4\Pgm$, bottom right: $2\Pgm 2\Pe$.
Two signal distributions are shown for \mlstar = 0.2 and 1\TeV, except the $2\Pgm 2\Pj$ channel where the trigger threshold
only allows searches for $\mlstar > 0.5\TeV$.
The asymmetric error bars
indicate the central confidence intervals for Poisson-distributed data and are obtained from the
Neyman construction as described in Ref.~\cite{Garwood1936}.}
 \label{fig:Mmin_muons}
\end{figure}

\subsection{L-shape search window}
\label{sec:selection-L}

After reconstruction of the intermediate boson (photon or $\PZ$-boson), two leptons remain to reconstruct the excited lepton,
either as $\ell\gamma$ or $\ell \PZ$.
Thus, both possible lepton+boson invariant masses are calculated, referred to in the following as $M_{min}^{X}$ and $M_{max}^{X}$ where $X$ is the channel considered, \estar or \mustar.
Figures~\ref{fig:Mmin_electrons} and \ref{fig:Mmin_muons} show  $M_{min}^{X}$ for all excited electron and muon channels with the background and systematic uncertainties described previously.

\begin{figure}[thbp]
\begin{center}
\includegraphics[width=0.49\textwidth]{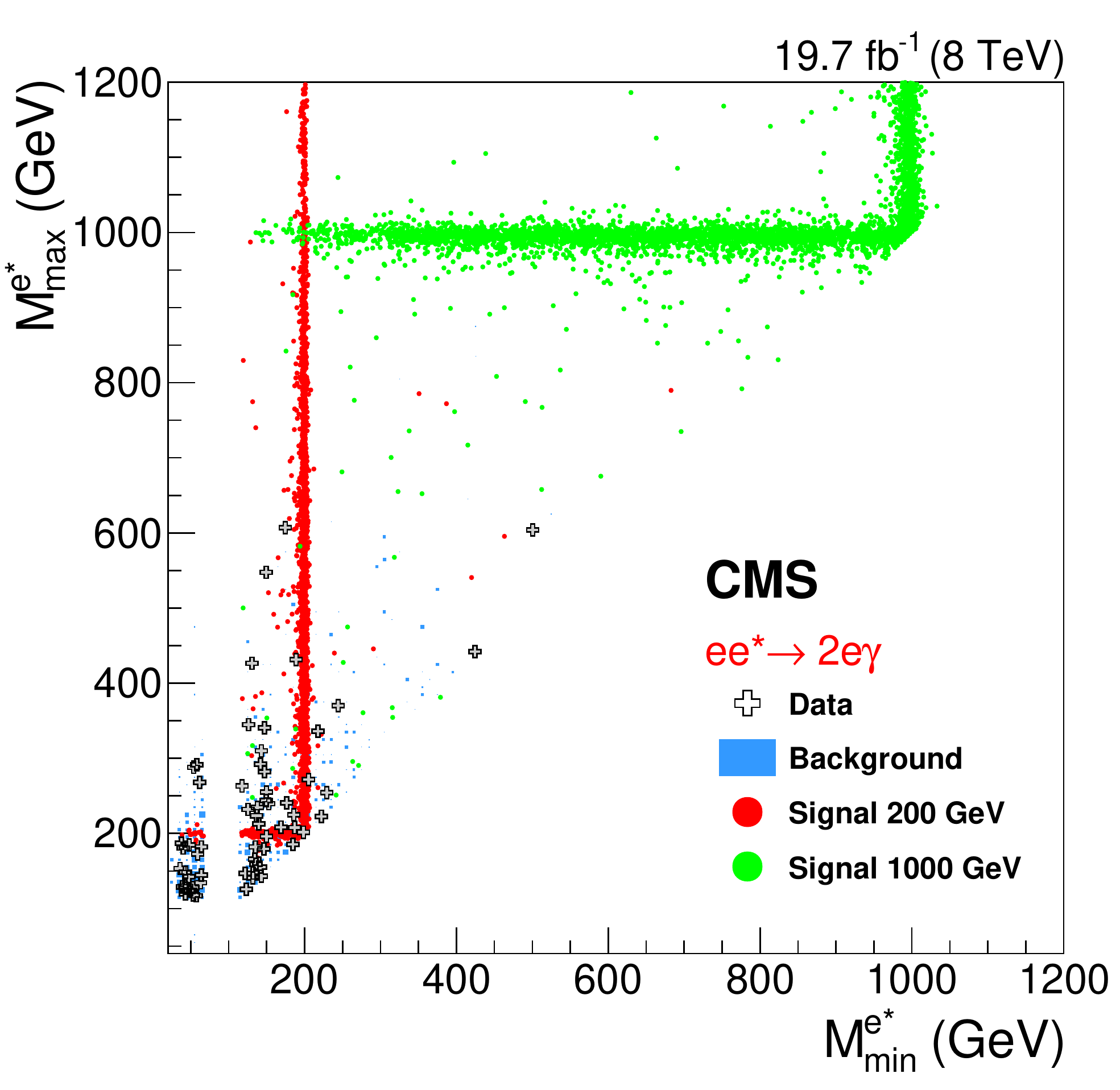}
\includegraphics[width=0.49\textwidth]{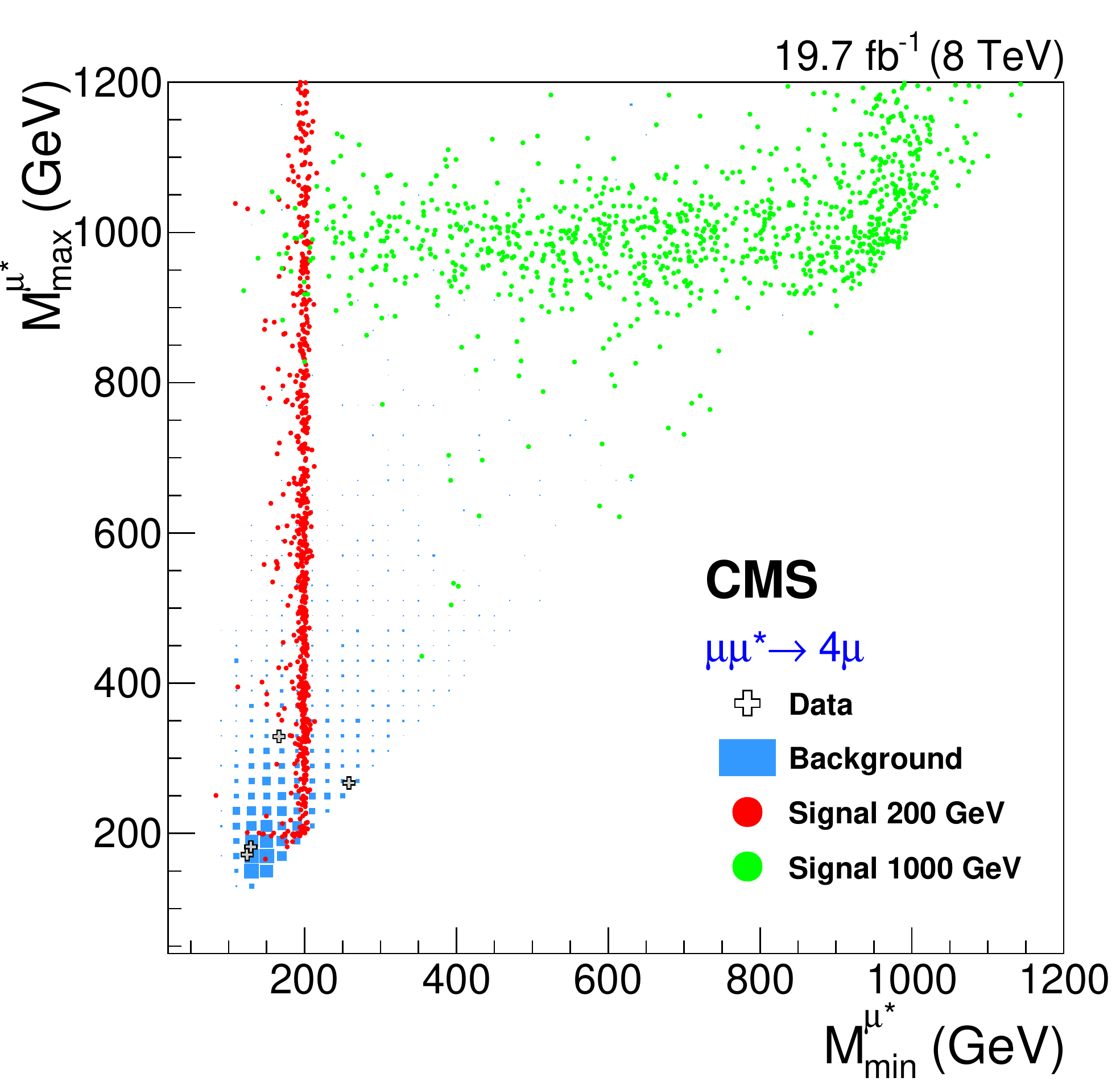}
\end{center}
\caption{Illustrative two dimensional minimal-versus-maximal invariant-mass distribution for the \eeg (left) and the $4\Pgm$ channel (right). It can be seen that the resolution worsens with increasing
signal mass and that the channels have different resolutions. The left plot clearly shows the effect of the additional $\PZ$-veto that is applied in this channel.
Background contributions are normalized to the given integrated luminosity,
while the normalization of the signal was
chosen to enhance visibility.
}
\label{fig:L_example}
\end{figure}

An illustrative plot of $M_{min}^{X}$ versus $M_{max}^{X}$ is given in Fig.~\ref{fig:L_example}. While the expected background tends to be at low invariant masses, a potential signal has the form of an inverted ``L'' around the excited lepton mass. Defining such a search window discriminates efficiently against background and is referred to in the following as the ``final selection'' or the ``L-shape cut'' when defining the final search regions.

\begin{figure}[thbp]
\begin{center}
\includegraphics[width=0.49\textwidth]{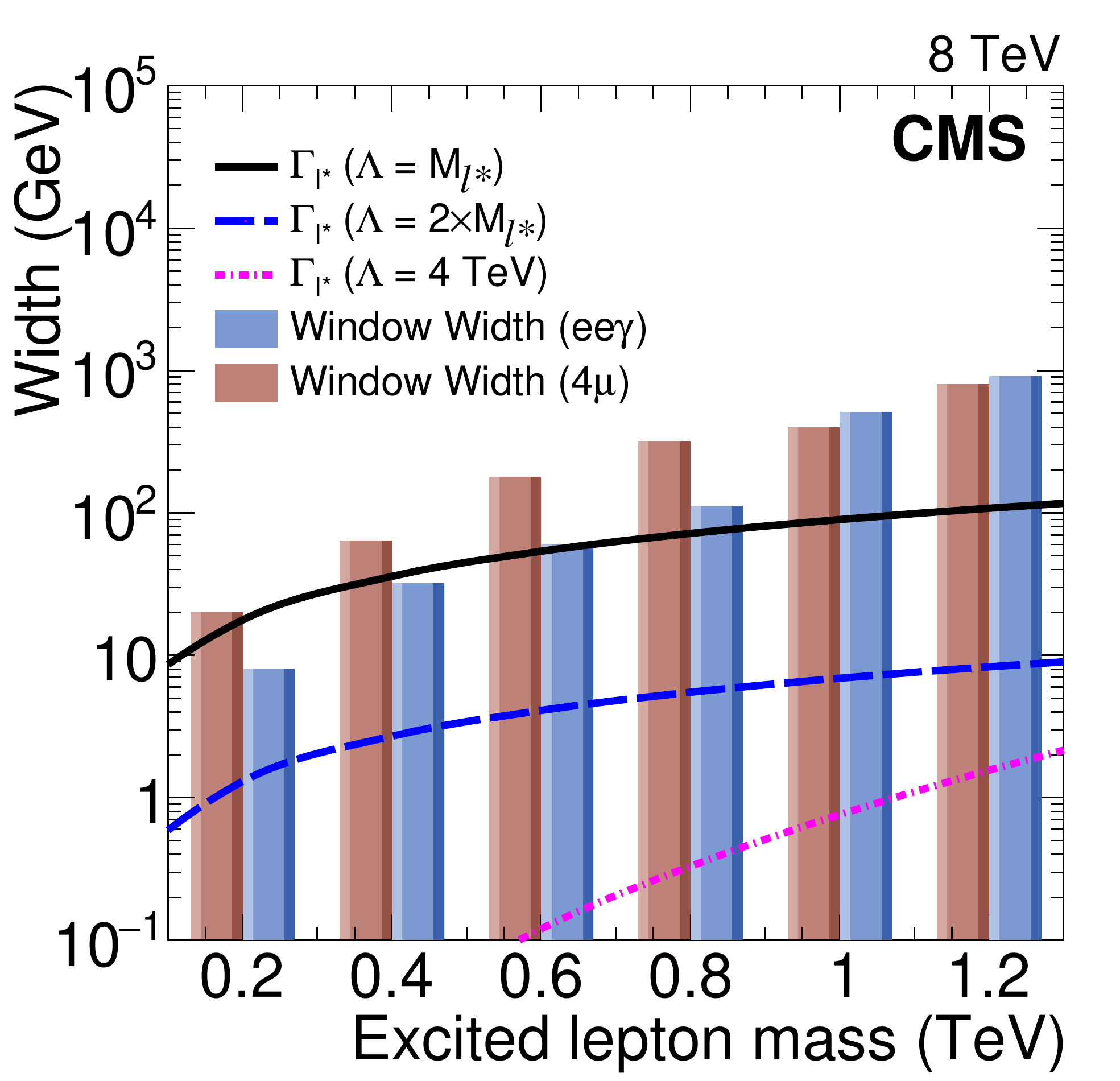}
\end{center}
\caption{
Width of the widest search window ($4\Pgm$ channel) and of a narrow one ($\Pe\Pe\gamma$) compared with
the intrinsic decay width of the excited lepton, as a function of the \lstar mass and for different
values of $\Lambda$. The latter shows the width of the excited leptons as defined in Ref.~\cite{Baur90},
including GM and CI decays, for the case $f = -f^{\prime} = 1$.} \label{fig:SearchWindows}
\end{figure}

The width of these L-shaped search regions depends on the channel and the \lstar mass. Detailed values for all channels are given in the Appendix. In the muon channels, the mass resolution worsens with increasing energy and the widths of the search windows need to become broader. This can be achieved without affecting the sensitivity of the search, since the high-mass regions are practically background-free. In the electron channels, the improving relative resolution of the electromagnetic calorimeter with increasing energy allows a more precise energy measurement at high masses. As a consequence, the width of the L-shaped windows is chosen individually for the different channels and mass points (by optimizing with respect to the best expected limit).
Shown in Fig.~\ref{fig:SearchWindows} is a comparison of the width of search window with the intrinsic excited lepton width as a function of the excited lepton mass, for representative values of the
compositeness scale $\Lambda$. This figure shows that the mass windows are in general much wider than the intrinsic width of the excited lepton, unless both its mass and $\Lambda$ are small. The size of the mass window has a negligible effect on the final result, as will be discussed in Section~\ref{sec:results}.

\begin{figure}[thbp]
\begin{center}
\includegraphics[width=0.49\textwidth]{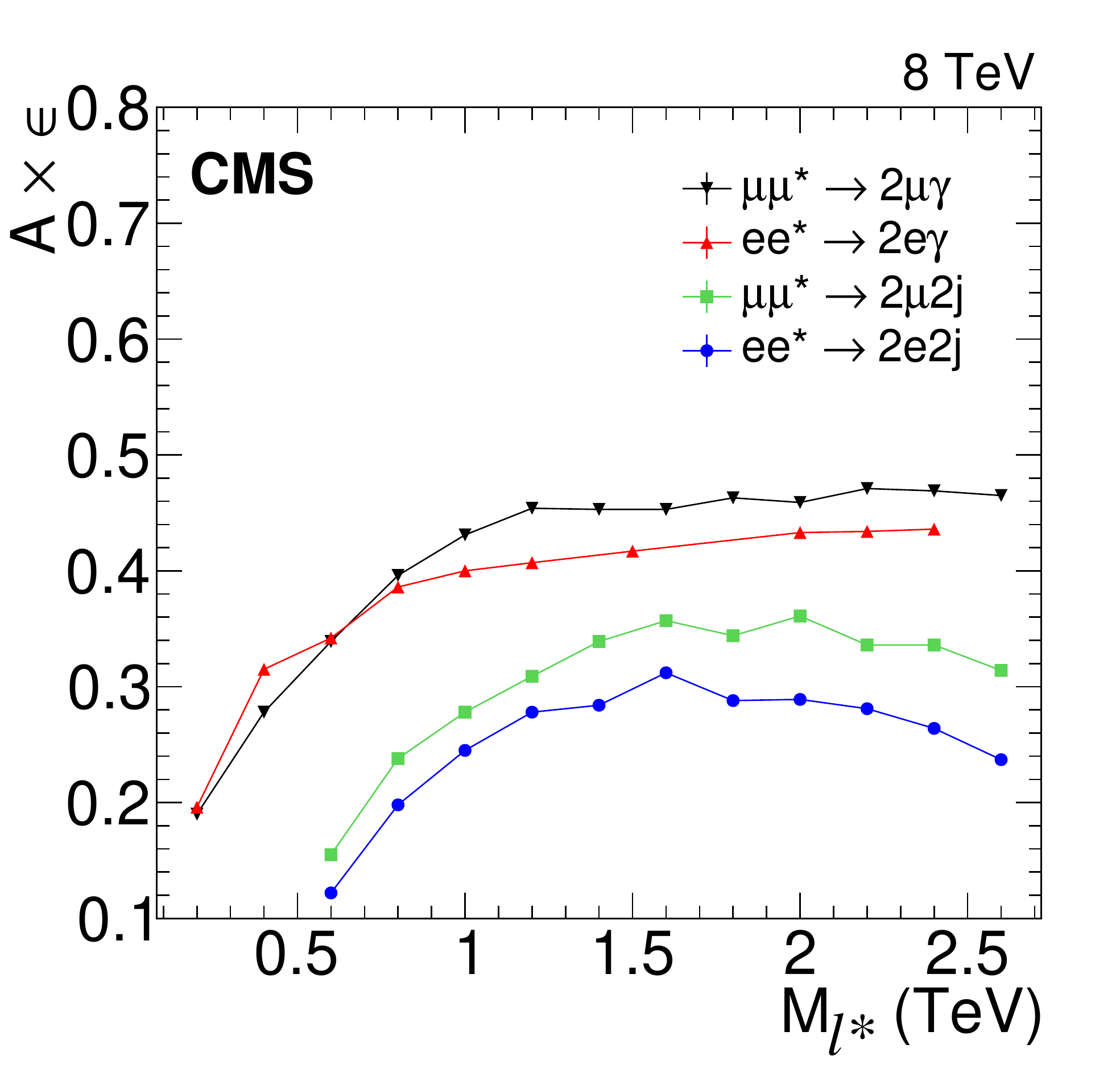}
\includegraphics[width=0.49\textwidth]{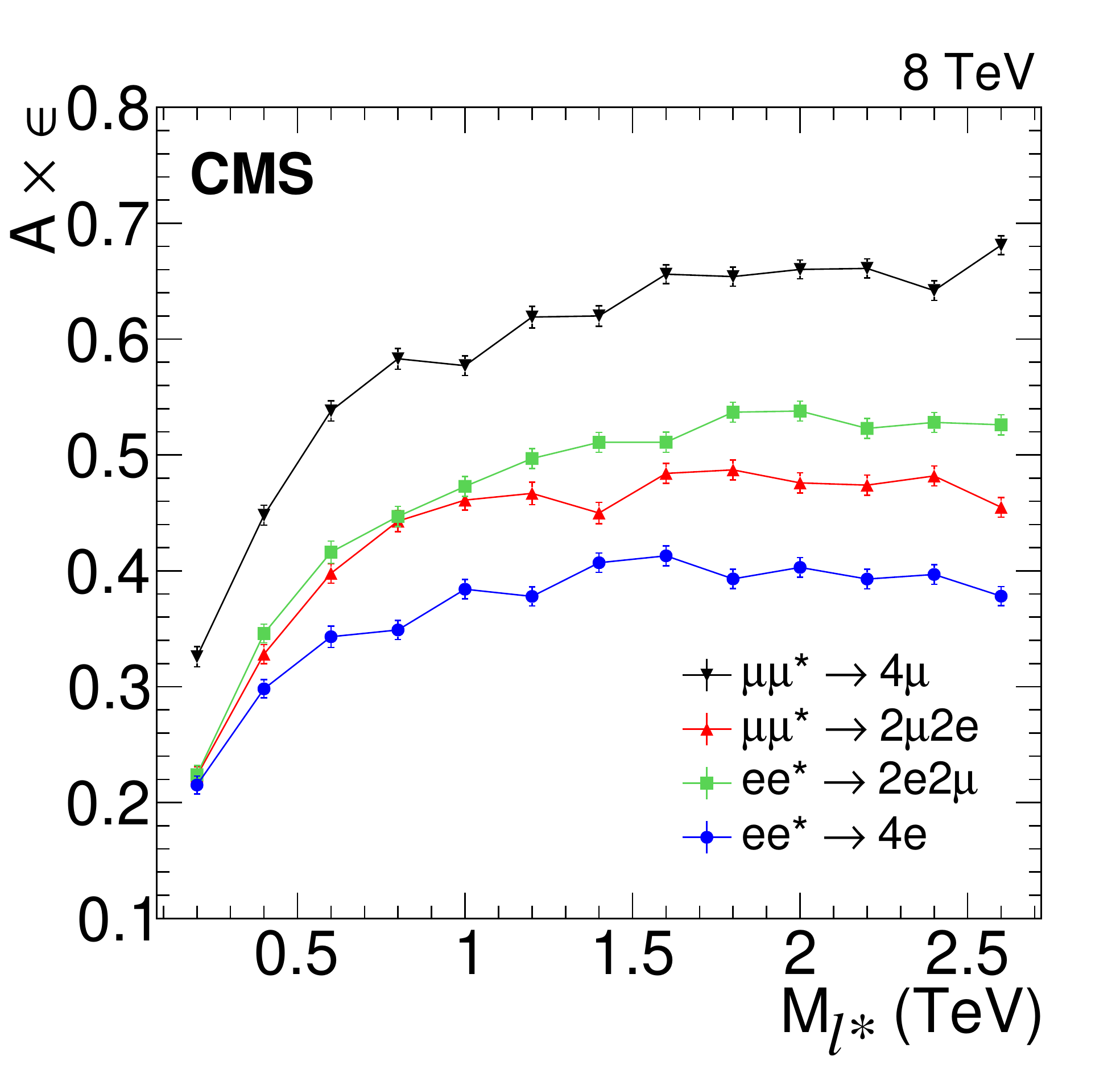}
\end{center}
\caption{The product of acceptance and efficiency after L-shape selection. In the left panel, the efficiencies for the \llg and $2\ell2\Pj$ channels are shown while the right panel gives the
efficiencies of the four-lepton channels. For the $2\ell2\Pj$ and $4\ell$ channels, the values do not include the branching fractions for the specific $\PZ$ boson decay channel.
}
\label{fig:eff}
\end{figure}

The product of acceptance and efficiency as a function of \lstar mass for all channels is shown in Fig.~\ref{fig:eff}.
The decreasing efficiency at high masses in the $2\ell2\Pj$ channels results from the subjettiness algorithm, which loses ability to resolve the
constituents of the
fat jets, which overlap more and more with increasing \lstar mass.

The selected L-shaped search regions with positions given by the simulated signal masses do not cover the complete \Mmin-\Mmax plane in the low mass region, where the search windows are narrow. To avoid
simulating more mass points, those regions are covered with additional L-shaped search regions based on a linear interpolation of the signal expectation between the two closest available simulated signal
masses. The $4\Pe$ channel is used to define the window positions that are adopted in all channels. There, the widths are estimated by linear interpolation between two consecutive masses such that the boundaries of all the
search regions are connected. The central positions of these resulting interpolated search windows are then applied in all channels, while the corresponding widths are estimated for each channel
individually.

The observed data, as well as the background expectation, in these newly defined L-shaped search regions
are
given by the \Mmin-\Mmax distributions. As there are no corresponding signal samples simulated for all these search regions
(cf. simulated signal samples as explained in Section ~\ref{sec:theo}), this information is not available for the signal.
The signal is therefore estimated by a fit to the signal expectation of the available simulated mass points including the
systematic uncertainties.

\subsection{Limits on cross section and compositeness scale \texorpdfstring{$\Lambda$}{Lambda}}
\label{sec:results}
\begin{figure}[hptb]
 \begin{center}
 \includegraphics[width=0.47\textwidth]{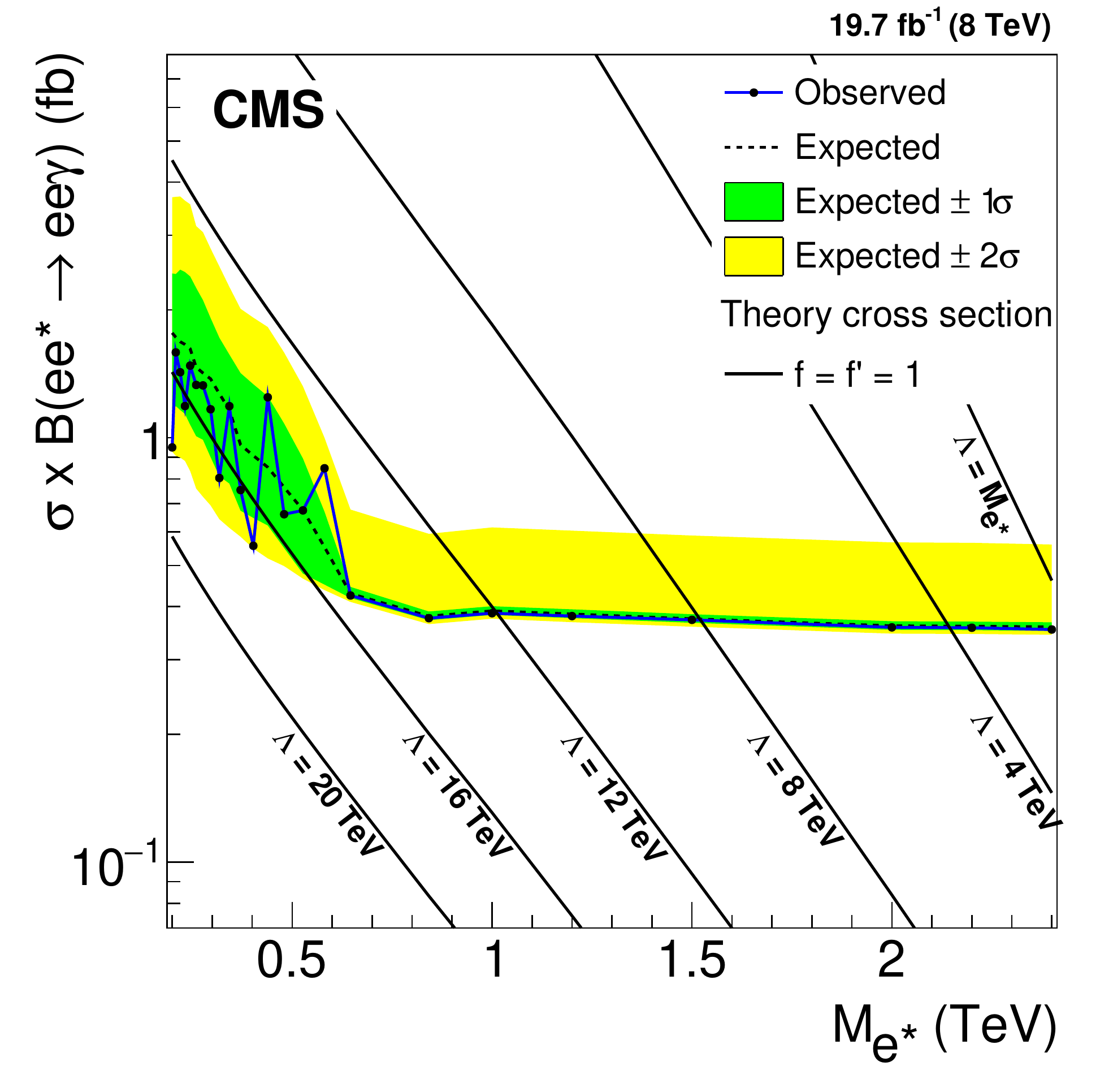}
 \includegraphics[width=0.47\textwidth]{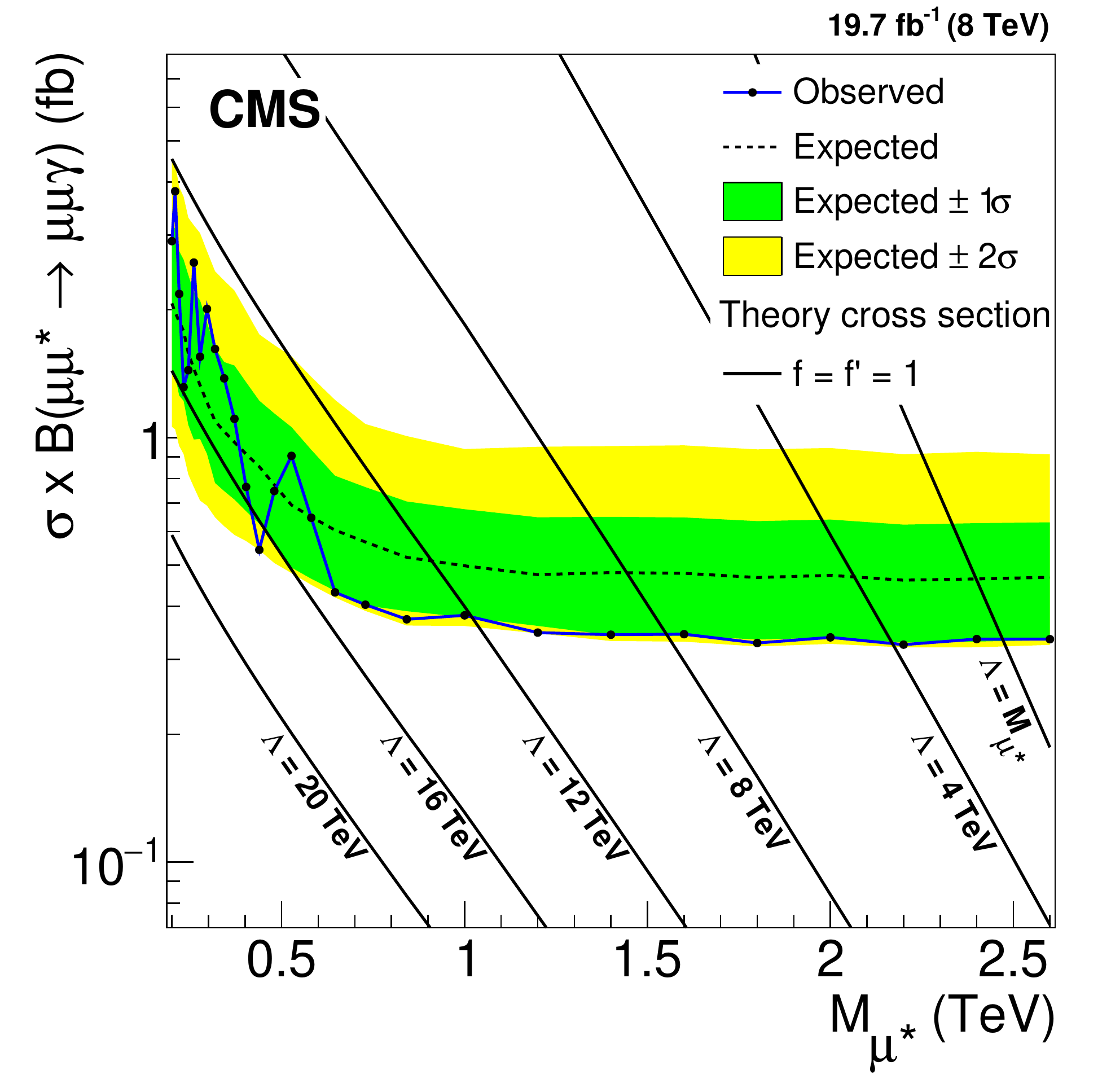}\\
 \includegraphics[width=0.47\textwidth]{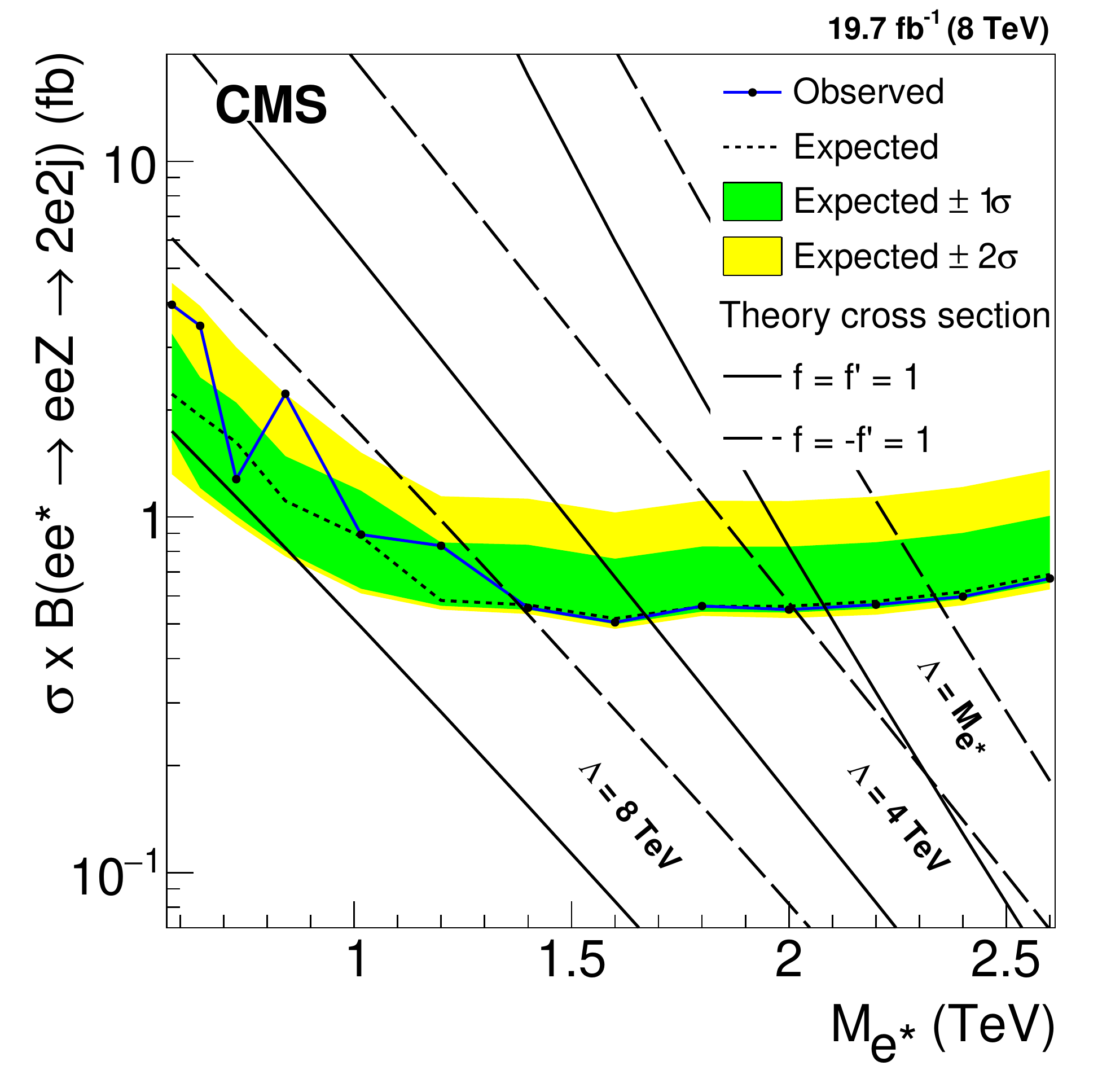}
 \includegraphics[width=0.47\textwidth]{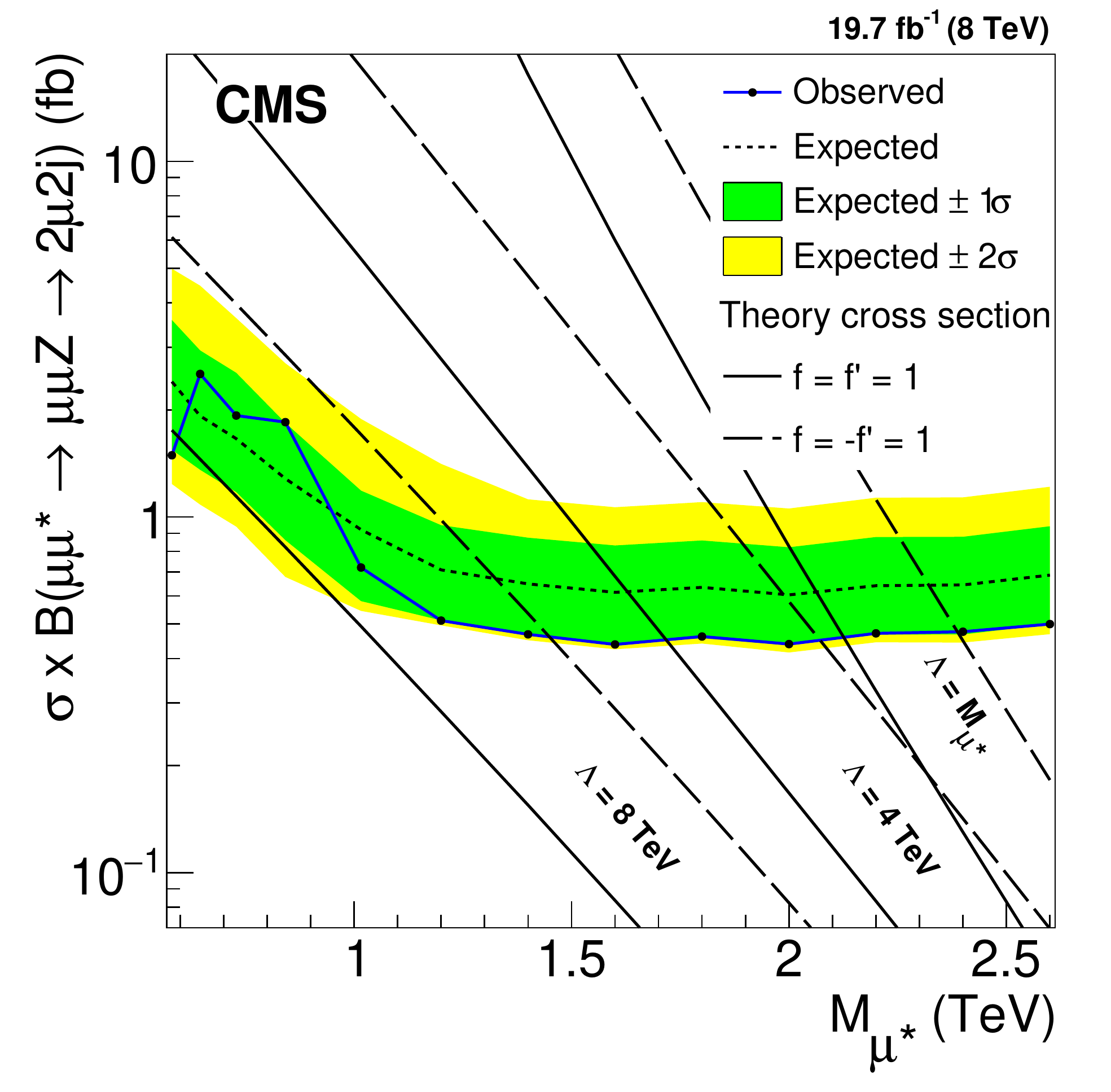} \\
 \includegraphics[width=0.47\textwidth]{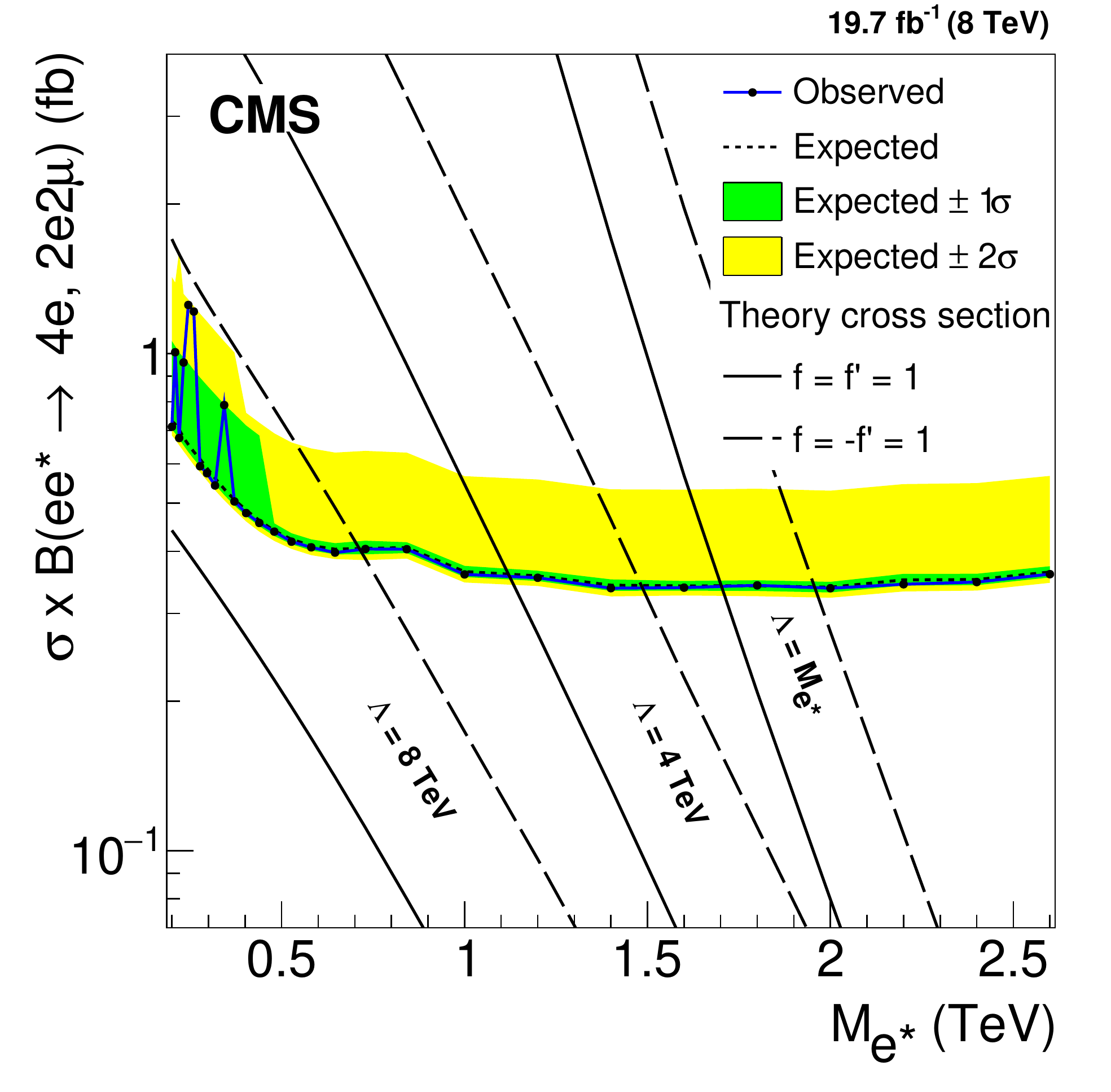}
 \includegraphics[width=0.47\textwidth]{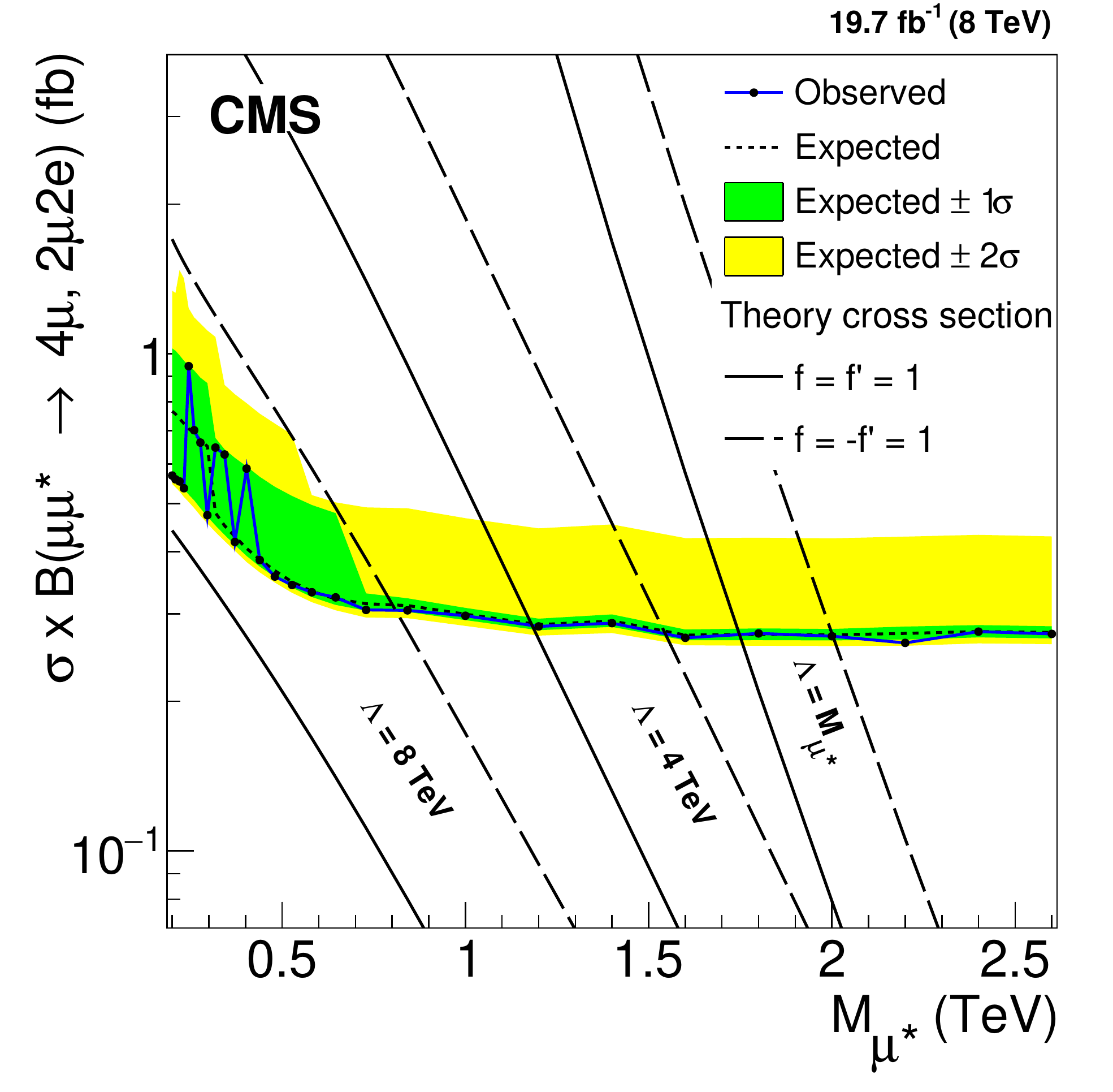}
 \end{center}
\caption{Upper limits at $95\%$ CL on the product of the production cross section and branching fraction for excited electrons (left) and excited muons (right). First row: \llg, second row: \lljj, last row:
combined four-lepton results. It is assumed that the signal efficiency is independent of $\Lambda$.
Theory curves are shown as solid or dashed lines.
}
\label{fig:xsec-limits}
\end{figure}

The resulting limits of cross section times braching fraction are shown in Fig.~\ref{fig:xsec-limits}. They range from  0.3\unit{fb} to 3\unit{fb} as a function of \mlstar. The four lepton final states: $4\Pe$ and $2\Pe2\Pgm$, $4\Pgm$ and $2\Pgm 2\Pe$, differing only in the decay of  the SM $\PZ$ boson, are combined. The other channels are shown individually. The black lines represent the theoretical cross sections including the NLO correction factors for different values of $\Lambda$. Solid lines are for the case $f = f^{\prime} = 1$ while the dashed lines are for $f = -f^{\prime} = 1$. The $95\%$ confidence level (CL) upper limit on the excited lepton production cross section times branching fraction has been set using a single-bin counting method \cite{higgstool}. The computation has been performed using a Bayesian~\cite{Bayes} approach.

The uncertainty bands have interesting behavior in some regions. They become asymmetric and in some cases the $1\sigma$ band disappears.
Both effects have their origin in the low background expectation in the corresponding search window. In such cases,
fluctuations of the limit to lower values are not possible.
Unstable behavior of both expected and observed limits is due to the limited number of (background) events in the search
regions, with the consequence that the presence of
a single event leads to a considerable upward fluctuation of the observed limit (see also tables in
the appendix).

\begin{figure}[thbp]
\begin{center}
\includegraphics[width=0.48\textwidth]{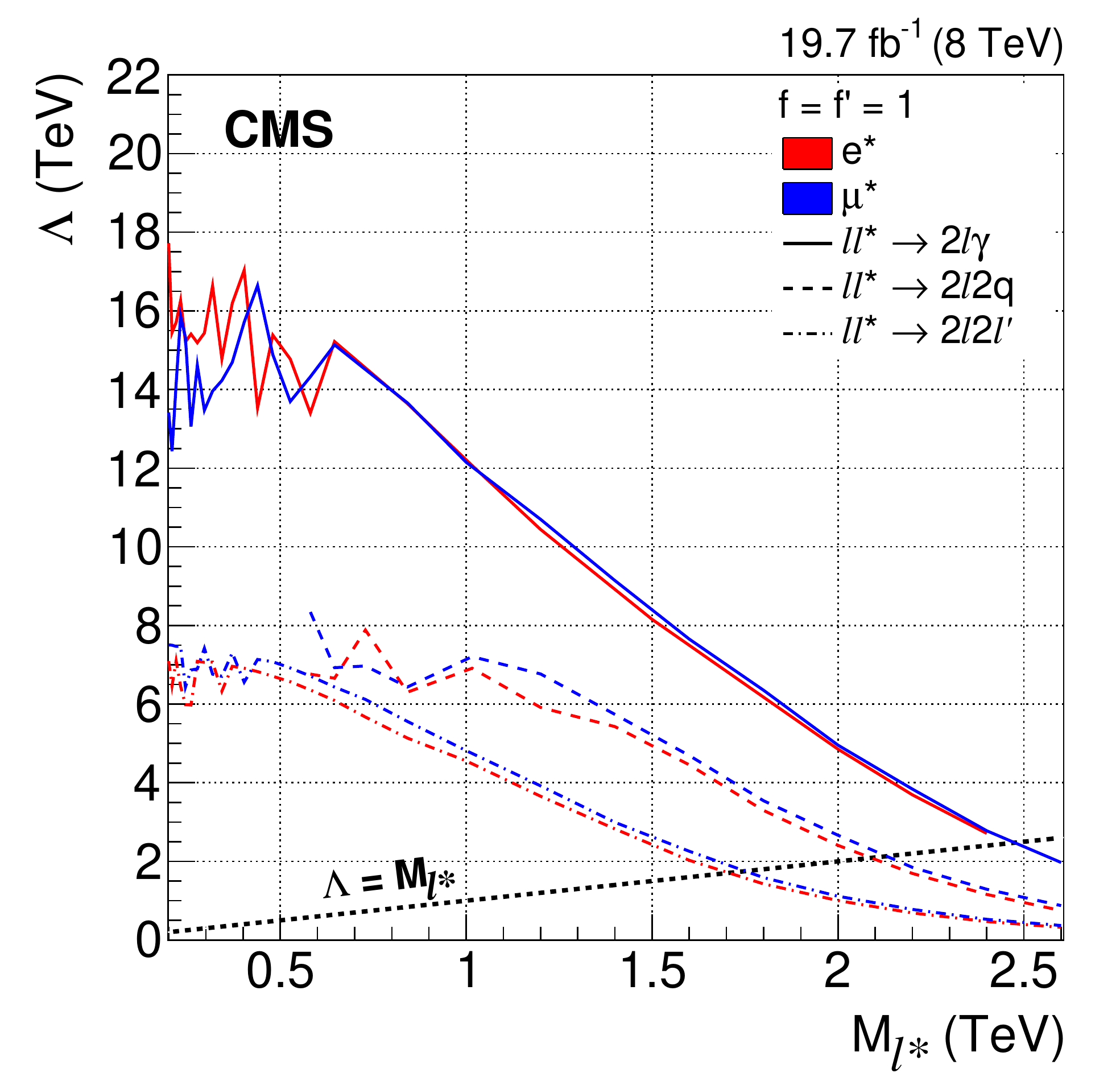}
\includegraphics[width=0.48\textwidth]{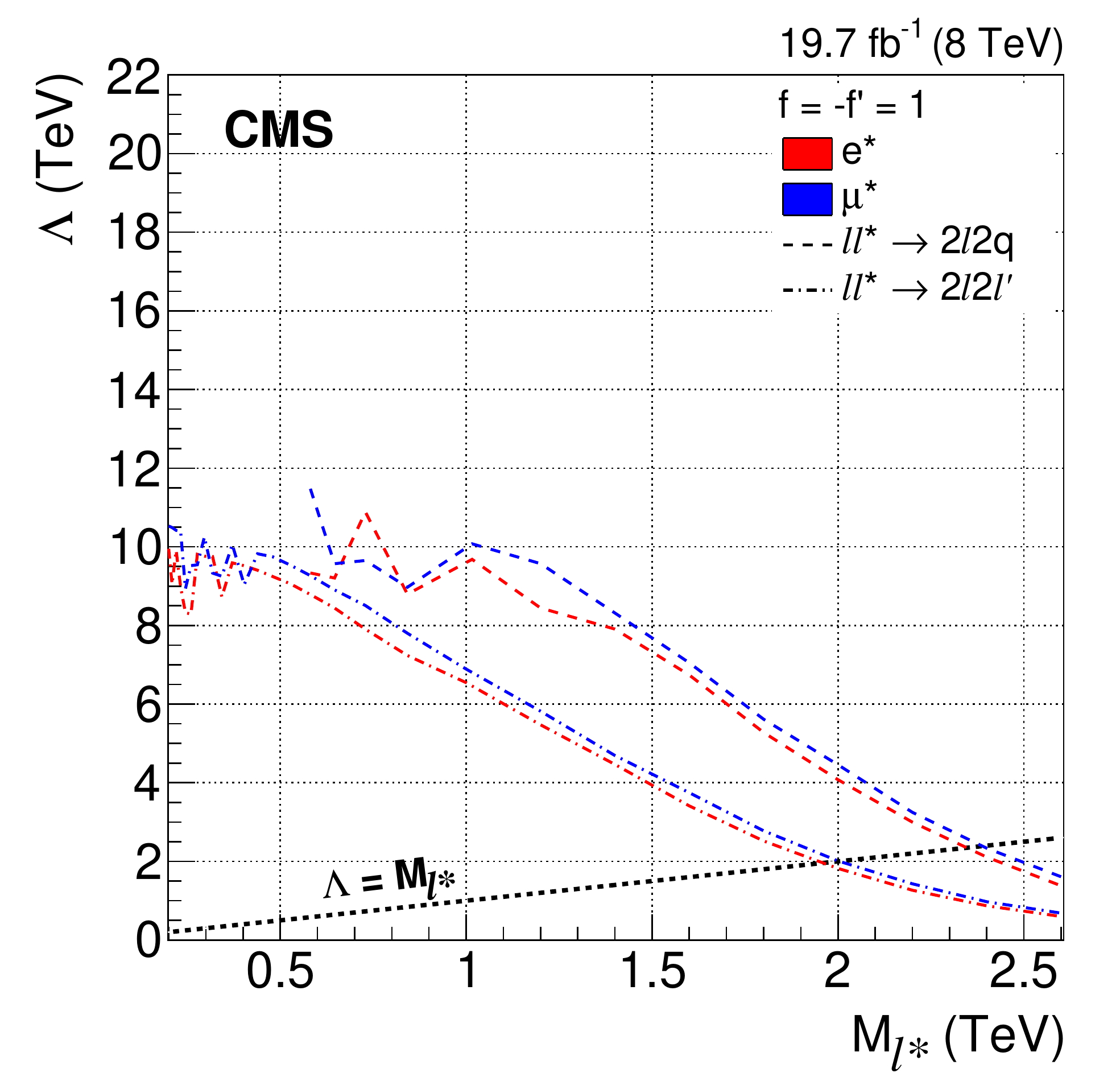}
\end{center}
\caption{Observed 95\% CL limits on the compositeness scale $\Lambda$
for the cases $f = f^{\prime} = 1$ and $f = -f^{\prime} = 1$,
as a function of the excited lepton mass for all channels. The excluded values are below the curves.}
\label{fig:LambdaComb}
\end{figure}

\begin{figure}[thbp]
\begin{center}
\includegraphics[width=0.8\textwidth]{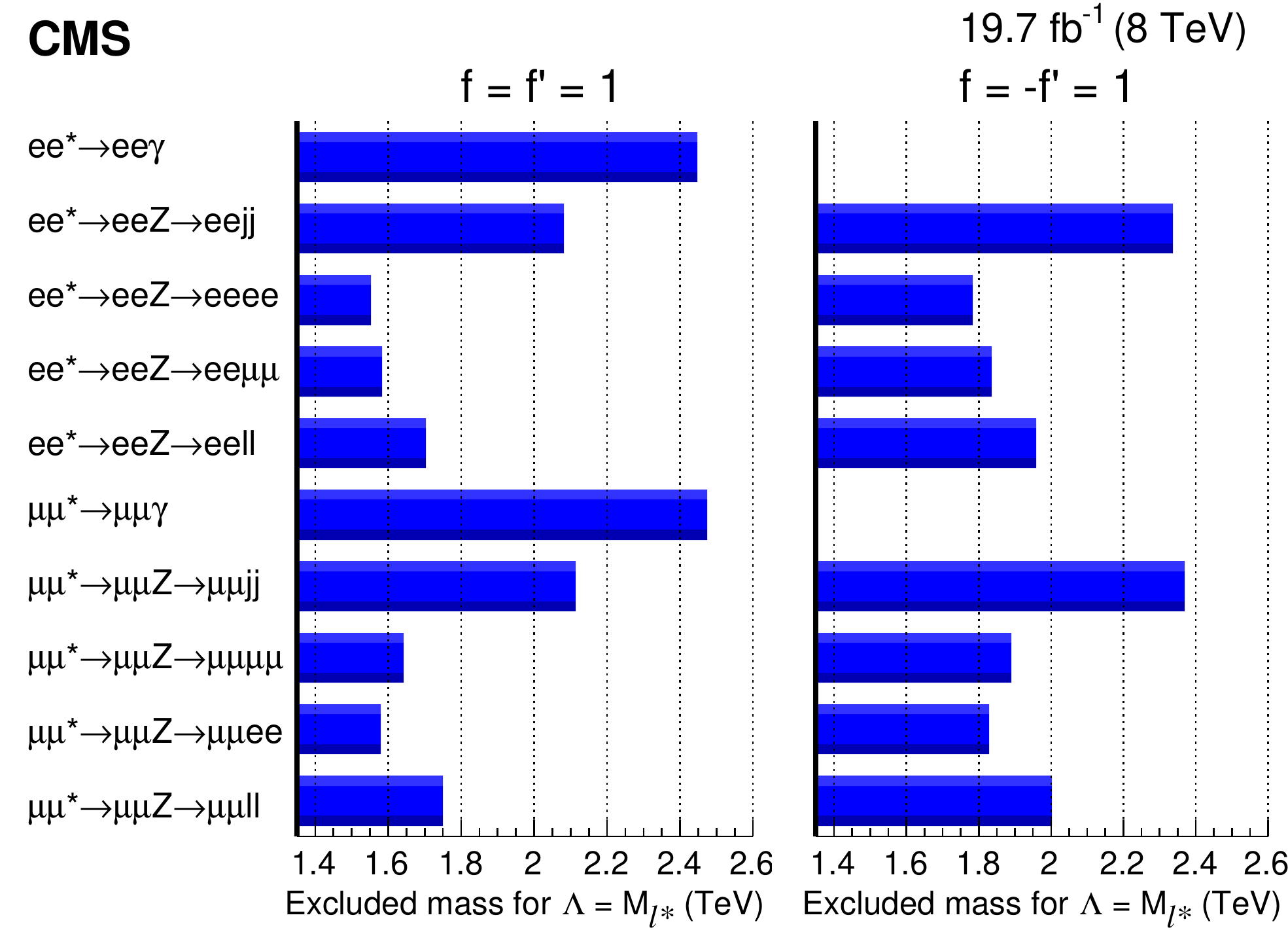}
\end{center}
\caption{Summary of all mass limits for the various channels and, including the combination of the four lepton channels,
for $M_{\ell^{*}} = \Lambda$.}
\label{fig:summary}
\end{figure}

The corresponding observed limits on the compositeness scale $\Lambda$ are displayed in Fig.~\ref{fig:LambdaComb}(left) for the case of SM-like couplings ($f = f^{\prime} = 1$) and in
Fig.~\ref{fig:LambdaComb}(right) for couplings of opposite sign ($f = -f^{\prime} = 1$). In the latter case, \llg cannot contribute. For low \mlstar masses compositeness scales up to 16\TeV can be excluded.
The sensitivity to $\Lambda$ decreases with increasing \mlstar. For the representative assumption of $\mlstar = \Lambda$, the resulting limits are summarized in Table~\ref{tab:SummaryLimits} and
Fig.~\ref{fig:summary}. Although, the assumption that the signal efficiency is independent of $\Lambda$ is not valid for the phase space where $\Lambda$ and $M_{\lstar}$ are small (lower left corner of
Figs.~\ref{fig:LambdaComb}), the strong $\Lambda$ dependence of the cross section, $\sigma \sim 1/\Lambda^{4}$, leads to a strong increase in sensitivity for low values of $\Lambda$ and $M_{\lstar}$ such that the
complete region under the limit curves in
Fig.~\ref{fig:LambdaComb} is nonetheless excluded.

Because of its considerably larger cross section times branching fraction, the \llg final state provides the maximum sensitivity for excluding excited leptons with masses up to 2.45\TeV. This limit
improves upon the existing ATLAS limit for single \lstar production based on  a partial 8\TeV data set \cite{atlas-limit_new} and exceeds significantly the limits of searches for single excited lepton
production at  previous colliders. The \llg channel shows no sensitivity for the case $f = -f^{\prime} = 1$, which is therefore studied with $\PZ$ boson radiation,
with the \lljj channel being dominant.
The  excited muon channels are slightly more sensitive
than
those of the excited electron channels, even though the resolution and thus the signal  separation ability of electron final states is higher than that of the muon channels. The higher exclusion power is due
to the better muon  reconstruction efficiency, which leads to an overall higher signal selection efficiency.

\begin{table}[hbtp]
\centering
\topcaption{Summary of the observed (expected) limits on \lstar mass, assuming \mlstar = $\Lambda$,
for the cases $f = f^{\prime} = 1$ and $f = -f^{\prime} = 1$. The latter case is not applicable to \llg.
}
\label{tab:SummaryLimits}
\begin{tabular}{l| c| c}
\hline
\multirow{2}{*}{Search channel }    & \multicolumn{2}{c}{ \mlstar $= \Lambda$, values in \TeV} \\
& $f = f^{\prime} = 1$                        & $f  = -f^{\prime} = 1$ \\  \hline
$ \Pe\Pe^* \, \, \to \Pe\Pe\gamma $        & 2.45 (2.45)    & \NA    \\
$\Pe\Pe^* \, \, \to \Pe\Pe\PZ  \, \to  2\Pe 2\Pj $       & 2.08 (2.07)    & 2.34 (2.33)   \\
$\Pe\Pe^* \, \, \to \Pe\Pe\PZ  \, \to  4\Pe  $         & 1.55 (1.55)    & 1.78 (1.78)   \\
$\Pe\Pe^* \, \, \to \Pe\Pe\PZ  \, \to  2\Pe 2\Pgm $         & 1.58 (1.58)    & 1.84 (1.84)   \\
$\Pe\Pe^* \, \, \to \Pe\Pe\PZ  \, \to  2\Pe 2\ell $         & 1.70 (1.70)    & 1.96 (1.96)   \\ \hline
$ \Pgm\Pgm^* \to \Pgm\Pgm\gamma$         & 2.47 (2.40)    & \NA    \\
$\Pgm\Pgm^* \to \Pgm\Pgm\PZ  \to 2\Pgm 2\Pj$              & 2.11 (2.05)    & 2.37 (2.31)   \\
$\mu\mu^*  \to \Pgm\Pgm\PZ  \to 4\Pgm$          & 1.64 (1.64)    & 1.89 (1.89)   \\
$\Pgm\Pgm^* \to \Pgm\Pgm\PZ  \to 2\Pgm 2\Pe$          & 1.58 (1.58)    & 1.83 (1.83)   \\
$\Pgm\Pgm^* \to \Pgm\Pgm\PZ  \to 2\Pgm 2\ell$         & 1.75 (1.75)    & 2.00 (2.00)   \\ \hline
\end{tabular}
\end{table}

\newpage
\section{Summary}

A comprehensive search for excited leptons, \estar and \mustar, in various channels has been performed using 19.7\fbinv of pp collision data at $\sqrt{s} = 8\TeV$. The excited lepton is assumed to be
produced via contact interactions in conjunction with the corresponding standard model lepton. Decaying to its ground state, the excited lepton may emit either a photon or a $\PZ$ boson.
No evidence of excited leptons is found
and exclusion limits are set on the compositeness scale $\Lambda$ as a function of the excited lepton mass \mlstar.

The \llg final state has the largest production cross section and has therefore previously been used for searches.
Following convention, the limits for the assumption $\Lambda = \mlstar$ are included here.
This final state yields the best limits, excluding excited electrons up to 2.45\TeV and excited muons up to 2.47\TeV, at 95\% confidence level.
These limits place the most stringent constraints to date on
the existence of excited leptons.

The $\lstar \to \ell \Z$ decay channel has been examined for the first time at hadron colliders, allowing the case where couplings between standard model leptons and excited
leptons $f = -f = 1$ can be studied. The leptonic
and hadronic (2-jet) final states of the $\PZ$ boson are used in this search; these final states are Lorentz boosted, requiring a dedicated reconstruction strategy.
The observed 95\% exclusion limits extend to 2.34 (2.37)\TeV for excited electrons (muons), for $f = -f^\prime = 1$.

\section*{Acknowledgements}
\enlargethispage{1cm}

\hyphenation{Bundes-ministerium Forschungs-gemeinschaft Forschungs-zentren} We congratulate our colleagues in the CERN accelerator departments for the excellent performance of the LHC and thank the technical and administrative staffs at CERN and at other CMS institutes for their contributions to the success of the CMS effort. In addition, we gratefully acknowledge the computing centres and personnel of the Worldwide LHC Computing Grid for delivering so effectively the computing infrastructure essential to our analyses. Finally, we acknowledge the enduring support for the construction and operation of the LHC and the CMS detector provided by the following funding agencies: the Austrian Federal Ministry of Science, Research and Economy and the Austrian Science Fund; the Belgian Fonds de la Recherche Scientifique, and Fonds voor Wetenschappelijk Onderzoek; the Brazilian Funding Agencies (CNPq, CAPES, FAPERJ, and FAPESP); the Bulgarian Ministry of Education and Science; CERN; the Chinese Academy of Sciences, Ministry of Science and Technology, and National Natural Science Foundation of China; the Colombian Funding Agency (COLCIENCIAS); the Croatian Ministry of Science, Education and Sport, and the Croatian Science Foundation; the Research Promotion Foundation, Cyprus; the Ministry of Education and Research, Estonian Research Council via IUT23-4 and IUT23-6 and European Regional Development Fund, Estonia; the Academy of Finland, Finnish Ministry of Education and Culture, and Helsinki Institute of Physics; the Institut National de Physique Nucl\'eaire et de Physique des Particules~/~CNRS, and Commissariat \`a l'\'Energie Atomique et aux \'Energies Alternatives~/~CEA, France; the Bundesministerium f\"ur Bildung und Forschung, Deutsche Forschungsgemeinschaft, and Helmholtz-Gemeinschaft Deutscher Forschungszentren, Germany; the General Secretariat for Research and Technology, Greece; the National Scientific Research Foundation, and National Innovation Office, Hungary; the Department of Atomic Energy and the Department of Science and Technology, India; the Institute for Studies in Theoretical Physics and Mathematics, Iran; the Science Foundation, Ireland; the Istituto Nazionale di Fisica Nucleare, Italy; the Ministry of Science, ICT and Future Planning, and National Research Foundation (NRF), Republic of Korea; the Lithuanian Academy of Sciences; the Ministry of Education, and University of Malaya (Malaysia); the Mexican Funding Agencies (CINVESTAV, CONACYT, SEP, and UASLP-FAI); the Ministry of Business, Innovation and Employment, New Zealand; the Pakistan Atomic Energy Commission; the Ministry of Science and Higher Education and the National Science Centre, Poland; the Funda\c{c}\~ao para a Ci\^encia e a Tecnologia, Portugal; JINR, Dubna; the Ministry of Education and Science of the Russian Federation, the Federal Agency of Atomic Energy of the Russian Federation, Russian Academy of Sciences, and the Russian Foundation for Basic Research; the Ministry of Education, Science and Technological Development of Serbia; the Secretar\'{\i}a de Estado de Investigaci\'on, Desarrollo e Innovaci\'on and Programa Consolider-Ingenio 2010, Spain; the Swiss Funding Agencies (ETH Board, ETH Zurich, PSI, SNF, UniZH, Canton Zurich, and SER); the Ministry of Science and Technology, Taipei; the Thailand Center of Excellence in Physics, the Institute for the Promotion of Teaching Science and Technology of Thailand, Special Task Force for Activating Research and the National Science and Technology Development Agency of Thailand; the Scientific and Technical Research Council of Turkey, and Turkish Atomic Energy Authority; the National Academy of Sciences of Ukraine, and State Fund for Fundamental Researches, Ukraine; the Science and Technology Facilities Council, UK; the US Department of Energy, and the US National Science Foundation.

Individuals have received support from the Marie-Curie programme and the European Research Council and EPLANET (European Union); the Leventis Foundation; the A. P. Sloan Foundation; the Alexander von Humboldt Foundation; the Belgian Federal Science Policy Office; the Fonds pour la Formation \`a la Recherche dans l'Industrie et dans l'Agriculture (FRIA-Belgium); the Agentschap voor Innovatie door Wetenschap en Technologie (IWT-Belgium); the Ministry of Education, Youth and Sports (MEYS) of the Czech Republic; the Council of Science and Industrial Research, India; the HOMING PLUS programme of the Foundation for Polish Science, cofinanced from European Union, Regional Development Fund; the OPUS programme of the National Science Center (Poland); the Compagnia di San Paolo (Torino); the Consorzio per la Fisica (Trieste); MIUR project 20108T4XTM (Italy); the Thalis and Aristeia programmes cofinanced by EU-ESF and the Greek NSRF; the National Priorities Research Program by Qatar National Research Fund; the Rachadapisek Sompot Fund for Postdoctoral Fellowship, Chulalongkorn University (Thailand); and the Welch Foundation, contract C-1845.

\newpage

\numberwithin{table}{section}

\section*{Appendix}
\label{sec:appendix}

Final numbers used to calculate the expected and observed cross section limits for the various excited lepton channels
are shown in Tab.~\ref{tab:A1}--~\ref{tab:A4}. In all tables, ``window'' refers to the interval between the upper and lower invariant mass boundaries of the search windows for the given mass
points. The interpolated search windows are not shown. The signal efficiency after all selection steps including the
search window is $\epsilon_\mathrm{signal}$. The expected number of events for the SM background and the number of observed data events
are $N_\mathrm{bg}$ and $N_\mathrm{data}$, respectively.

\begin{table}[hp!]
\renewcommand{\arraystretch}{1.2}
\begin{center}
{
\topcaption{Final numbers used to calculate the cross section limits for the excited lepton channels resulting in photon emission, \mmg and \eeg. \label{tab:A1}
}
\resizebox{\textwidth}{!}{
\begin{tabular}{c|c|ccc|c|ccc}
\hline
$M_{\ell^{*}}$ & \multicolumn{4}{c|}{\mmg \phantom{xxxxx} } & \multicolumn{4}{c}{\eeg \phantom{xxxxx} } \\
(GeV) & Window (GeV) & $\epsilon_\mathrm{signal}$ & $N_\mathrm{bg}$ & $N_\mathrm{data}$ & Window (\GeVns) & $\epsilon_\mathrm{signal}$ & $N_\mathrm{bg}$ & $N_\mathrm{data}$ \\
\hline
200 & 194--206\phantom{0}  & 19.0\% & $6.95\pm 1.64$ & 10& 196--204\phantom{0}  & 19.6\% & $4.57\pm 1.21$ & 1 \\
400 &  384--416\phantom{0}  & 27.8\% & $1.27\pm 0.60$ & 1 & 384--416\phantom{0}  & 31.5\% & $1.19\pm 0.61$ & 0 \\
600 &  564--636\phantom{0}  & 33.9\% & $0.64\pm 0.48$ & 0 & 570--630\phantom{0}  & 34.2\% & $0.40\pm 0.31$ & 2 \\
800 &  720--880\phantom{0}  & 39.6\% & $0.29\pm 0.28$ & 0 & 744--856\phantom{0}  & 38.6\% & $0.01\pm 0.01$ & 0 \\
1000 &  720--1280 & 43.1\% & $0.29\pm 0.28$ & 0 & 744--1256 & 40.0\% & $0.05\pm 0.04$ & 0 \\
1200 &  720--1680 & 45.4\% & $0.57\pm 0.40$ & 0 & 744--1656 & 40.7\% & $0.05\pm 0.04$ & 0 \\
1400 &  720--2080 & 45.3\% & $0.57\pm 0.40$ & 0 &\NA       &\NA     &\NA             &\NA\\
1500 &\NA       &\NA     &\NA             &\NA& 744--2256 & 41.7\% & $0.05\pm 0.04$ & 0 \\
1600 &  720--2480 & 45.3\% & $0.57\pm 0.40$ & 0 &\NA       &\NA     &\NA             &\NA\\
1800 &  720--2880 & 46.3\% & $0.57\pm 0.40$ & 0 &\NA       &\NA     &\NA             &\NA\\
2000 &  720--3280 & 45.9\% & $0.57\pm 0.40$ & 0 & 744--3256 & 43.3\% & $0.05\pm 0.04$ & 0 \\
2200 &  720--3680 & 47.1\% & $0.57\pm 0.40$ & 0 & 744--3656 & 43.4\% & $0.05\pm 0.04$ & 0 \\
2400 &  720--4080 & 46.9\% & $0.57\pm 0.40$ & 0 & 744--4056 & 43.6\% & $0.05\pm 0.04$ & 0 \\
2600 &  720--4480 & 46.5\% & $0.57\pm 0.40$ & 0 &\NA       &\NA     &\NA             &\NA\\
\hline
\end{tabular}
}
}

\vspace{1cm}

{
\topcaption{Final numbers used to calculate the cross section limits for the excited lepton channels
resulting in the emission of a $\PZ$ boson that decays to two jets, \mmjj and \eejj. \label{tab:A2}
}
\resizebox{\textwidth}{!}{
\begin{tabular}{c|c|ccc|c|ccc}
\hline
$M_{\ell^{*}}$ & \multicolumn{4}{c|}{\mmjj} & \multicolumn{4}{c}{\eejj} \\
(GeV) & Window (GeV) & $\epsilon_\mathrm{signal}$ & $N_\mathrm{bg}$ & $N_\mathrm{data}$ & Window (GeV) & $\epsilon_\mathrm{signal}$ & $N_\mathrm{bg}$ & $N_\mathrm{data}$ \\
\hline
600 & \phantom{0}558--642\phantom{0} & 15.5\% & $4.69\pm 1.58$ & 3 & \phantom{0}570--630\phantom{0} & 12.2\% & $3.19\pm 1.11$ & 3 \\
800 & \phantom{0}728--872\phantom{0} & 23.8\% & $3.35\pm 1.15$ & 4 & \phantom{0}728--856\phantom{0} & 19.8\% & $2.49\pm 0.88$ & 3 \\
1000 & \phantom{0}900--1100 & 27.8\% & $1.75\pm 0.63$ & 1 & 900--1100 & 24.5\% & $1.47\pm 0.55$ & 1 \\
1200 & 1068--1332 & 30.9\% & $0.94\pm 0.37$ & 0 & 1068--1332 & 27.8\% & $0.50\pm 0.26$ & 1 \\
1400 & 1100--1700 & 33.9\% & $0.70\pm 0.30$ & 0 & 1200--1600 & 28.4\% & $0.50\pm 0.23$ & 0 \\
1600 & 1100--2100 & 35.7\% & $0.70\pm 0.30$ & 0 & 1200--2000 & 31.2\% & $0.50\pm 0.23$ & 0 \\
1800 & 1100--2500 & 34.4\% & $0.70\pm 0.30$ & 0 & 1200--2400 & 28.8\% & $0.50\pm 0.23$ & 0 \\
2000 & 1100--2900 & 36.1\% & $0.70\pm 0.30$ & 0 & 1200--2800 & 28.9\% & $0.50\pm 0.23$ & 0 \\
2200 & 1100--3300 & 33.6\% & $0.70\pm 0.30$ & 0 & 1200--3200 & 28.1\% & $0.50\pm 0.23$ & 0 \\
2400 & 1100--3700 & 33.6\% & $0.70\pm 0.30$ & 0 & 1200--3600 & 26.4\% & $0.50\pm 0.23$ & 0 \\
2600 & 1100--4100 & 31.4\% & $0.70\pm 0.30$ & 0 & 1200--4000 & 23.7\% & $0.50\pm 0.23$ & 0 \\
\hline
\end{tabular}
}
}
\end{center}
\end{table}

\begin{table}[hp!]
\renewcommand{\arraystretch}{1.2}
\begin{center}
{
\topcaption{Final numbers used to calculate the cross section limits for the two excited muon channels in the 4$\ell$ final states. \label{tab:A3}
}
\resizebox{\textwidth}{!}{
\begin{tabular}{c|c|ccc|c|ccc}
\hline
$M_{\Pgm^{*}}$ & \multicolumn{4}{c|}{\mmmm} & \multicolumn{4}{c}{\mmee} \\
(GeV) & Window (GeV) & $\epsilon_\mathrm{signal}$ & $N_\mathrm{bg}$ & $N_\mathrm{data}$ & Window (GeV) & $\epsilon_\mathrm{signal}$ & $N_\mathrm{bg}$ & $N_\mathrm{data}$ \\
\hline
200 & 190--210\phantom{0} & 32.6\% & $0.77\pm 0.12$ & 0 & \phantom{0}196--204\phantom{0} & 22.3\% & $0.23\pm 0.05$ & 0 \\
400 & 368--432\phantom{0} & 44.8\% & $0.23\pm 0.04$ & 0 & \phantom{0}376--424\phantom{0} & 32.8\% & $0.14\pm 0.03$ & 1 \\
600 & 510--690\phantom{0} & 53.8\% & $0.13\pm 0.02$ & 0 & \phantom{0}540--660\phantom{0} & 39.8\% & $0.07\pm 0.03$ & 0 \\
800 & 640--960\phantom{0} & 58.3\% & $0.06\pm 0.01$ & 0 & \phantom{0}720--880\phantom{0} & 44.3\% & $0.04\pm 0.01$ & 0 \\
1000 & 800--1200 & 57.7\% & $0.03\pm 0.01$ & 0 & \phantom{0}850--1150 & 46.1\% & $0.01\pm 0.01$ & 0 \\
1200 & 800--1600 & 61.9\% & $0.04\pm 0.01$ & 0 & 1000--1400 & 46.7\% & $0.00\pm 0.00$ & 0 \\
1400 & 800--2000 & 62.0\% & $0.04\pm 0.01$ & 0 & 1200--1800 & 45.0\% & $0.01\pm 0.01$ & 0 \\
1600 & 800--2400 & 65.6\% & $0.04\pm 0.01$ & 0 & 1200--2200 & 48.4\% & $0.01\pm 0.01$ & 0 \\
1800 & 800--2800 & 65.4\% & $0.04\pm 0.01$ & 0 & 1200--2600 & 48.7\% & $0.01\pm 0.01$ & 0 \\
2000 & 800--3200 & 66.0\% & $0.04\pm 0.01$ & 0 & 1200--3000 & 47.6\% & $0.01\pm 0.01$ & 0 \\
2200 & 800--3600 & 66.1\% & $0.04\pm 0.01$ & 0 & 1200--3400 & 47.4\% & $0.01\pm 0.01$ & 0 \\
2400 & 800--4000 & 64.2\% & $0.04\pm 0.01$ & 0 & 1200--3800 & 48.2\% & $0.01\pm 0.01$ & 0 \\
2600 & 800--4400 & 68.1\% & $0.04\pm 0.01$ & 0 & 1200--4200 & 45.5\% & $0.01\pm 0.01$ & 0 \\
\hline
\end{tabular}
}
}

\vspace{1cm}

{
\topcaption{Final numbers used to calculate the cross section limits for the two excited electron channels in the 4$\ell$ final states. \label{tab:A4}
}
\resizebox{\textwidth}{!}{
\begin{tabular}{c|c|ccc|c|ccc}
\hline
$M_{\Pe^{*}}$ & \multicolumn{4}{c|}{\eeee} & \multicolumn{4}{c}{\eemm} \\
(GeV) & Window (GeV) & $\epsilon_\mathrm{signal}$ & $N_\mathrm{bg}$ & $N_\mathrm{data}$ & Window (GeV) & $\epsilon_\mathrm{signal}$ & $N_\mathrm{bg}$ & $N_\mathrm{data}$ \\
\hline
200 & \phantom{0}196--204\phantom{0} & 21.5\% & $0.23\pm 0.05$ & 0 & 196--204\phantom{0} & 22.4\% & $0.24\pm 0.05$ & 0 \\
400 & \phantom{0}384--416\phantom{0} & 29.8\% & $0.08\pm 0.02$ & 0 & 384--416\phantom{0} & 34.6\% & $0.09\pm 0.02$ & 0 \\
600 & \phantom{0}570--630\phantom{0} & 34.3\% & $0.03\pm 0.01$ & 0 & 552--648\phantom{0} & 41.6\% & $0.08\pm 0.02$ & 0 \\
800 & \phantom{0}744--856\phantom{0} & 34.9\% & $0.01\pm 0.00$ & 0 & 728--872\phantom{0} & 44.7\% & $0.02\pm 0.01$ & 0 \\
1000 & \phantom{0}900--1100 & 38.4\% & $0.01\pm 0.00$ & 0 & 860--1140 & 47.3\% & $0.02\pm 0.01$ & 0 \\
1200 & 1000--1200 & 37.8\% & $0.01\pm 0.01$ & 0 & 860--1540 & 49.7\% & $0.02\pm 0.01$ & 0 \\
1400 & 1000--1600 & 40.7\% & $0.01\pm 0.01$ & 0 & 860--1940 & 51.1\% & $0.02\pm 0.01$ & 0 \\
1600 & 1000--2000 & 41.3\% & $0.01\pm 0.01$ & 0 & 860--2340 & 51.1\% & $0.02\pm 0.01$ & 0 \\
1800 & 1000--2400 & 39.3\% & $0.01\pm 0.01$ & 0 & 860--2740 & 53.7\% & $0.02\pm 0.01$ & 0 \\
2000 & 1000--2800 & 40.3\% & $0.01\pm 0.01$ & 0 & 860--3140 & 53.8\% & $0.02\pm 0.01$ & 0 \\
2200 & 1000--3200 & 39.3\% & $0.01\pm 0.01$ & 0 & 860--3540 & 52.3\% & $0.02\pm 0.01$ & 0 \\
2400 & 1000--3800 & 39.7\% & $0.01\pm 0.01$ & 0 & 860--3940 & 52.8\% & $0.02\pm 0.01$ & 0 \\
2600 & 1000--4200 & 37.8\% & $0.01\pm 0.01$ & 0 & 860--4340 & 52.6\% & $0.02\pm 0.01$ & 0 \\
\hline
\end{tabular}
}
}
\end{center}
\end{table}

\newpage
\bibliography{auto_generated}

\cleardoublepage \section{The CMS Collaboration \label{app:collab}}\begin{sloppypar}\hyphenpenalty=5000\widowpenalty=500\clubpenalty=5000\textbf{Yerevan Physics Institute,  Yerevan,  Armenia}\\*[0pt]
V.~Khachatryan, A.M.~Sirunyan, A.~Tumasyan
\vskip\cmsinstskip
\textbf{Institut f\"{u}r Hochenergiephysik der OeAW,  Wien,  Austria}\\*[0pt]
W.~Adam, E.~Asilar, T.~Bergauer, J.~Brandstetter, E.~Brondolin, M.~Dragicevic, J.~Er\"{o}, M.~Flechl, M.~Friedl, R.~Fr\"{u}hwirth\cmsAuthorMark{1}, V.M.~Ghete, C.~Hartl, N.~H\"{o}rmann, J.~Hrubec, M.~Jeitler\cmsAuthorMark{1}, V.~Kn\"{u}nz, A.~K\"{o}nig, M.~Krammer\cmsAuthorMark{1}, I.~Kr\"{a}tschmer, D.~Liko, T.~Matsushita, I.~Mikulec, D.~Rabady\cmsAuthorMark{2}, B.~Rahbaran, H.~Rohringer, J.~Schieck\cmsAuthorMark{1}, R.~Sch\"{o}fbeck, J.~Strauss, W.~Treberer-Treberspurg, W.~Waltenberger, C.-E.~Wulz\cmsAuthorMark{1}
% ...
\vskip\cmsinstskip
% ...
1:~~Also at Vienna University of Technology, Vienna, Austria\\
2:~~Also at CERN, European Organization for Nuclear Research, Geneva, Switzerland\\
% ...

\end{sloppypar}
\end{document}